\documentstyle[preprint,aps,epsfig]{revtex}
\tighten
\draft
\title{Effective Weak Chiral Lagrangian to 
${\cal O} (p^4)$ \\ in the Chiral Quark Model
\footnote{Final version for publication in {\em Nucl. Phys}. {\bf B}.}} 
\preprint{\vbox{\hbox{hep-ph/9903275} 
\hbox{RUB-TPII-10/98} \hbox{PNU-NTG-01/98}}}
\author{
Mario Franz$^{(1)}$
\footnote{email:mariof@tp2.ruhr-uni-bochum.de},
Hyun-Chul Kim$^{(2)}$
\footnote{email:hchkim@hyowon.pusan.ac.kr},
and Klaus Goeke$^{(1,3)}$
\footnote{email:klaus.goeke@ruhr-uni-bochum.de}
} 
\address{(1)
Institute for Theoretical  Physics  II,   P.O. Box 102148,  \\
Ruhr-University Bochum,
 D-44780 Bochum,   Germany \\
(2) Department of Physics, Pusan National University,\\
609-735 Pusan, Republic of Korea,\\
(3) RCNP, University of Osaka, Osaka, Japan}
\date{August, 1999}
\begin{document}
\bibliographystyle{prsty}
\maketitle
\begin{abstract}
We investigate the $\Delta S = 1,2$ effective weak chiral Lagrangian
within the framework of the chiral quark model.  Starting from the
effective four-quark operators, we derive the effective weak chiral 
action by integrating out the constituent quark fields.  Employing 
the derivative expansion, we obtain the effective weak chiral Lagrangian 
to order ${\cal O} (p^4)$.  We examine the contributions of the order 
${\cal O} (N_c)$ to the ratio $g_{\underline{8}}/g_{\underline{27}}$
, considering e.g. the quark axial-vector constant $g_{\rm A}$ different 
from unity.  The low energy 
constants of the counterterms are also presented and discussed.    
\vspace{1cm}

\end{abstract}
\date{January, 1999}

\section{Introduction}
Processes involving the creation or annihilation of strangeness are described 
in the Standard Model by $W$-exchange.  While the theoretical formulation is 
simple at scales around the $W$-mass the description of nonleptonic decays 
of light hadrons at low energies is complicated and difficult because of 
the presence of the strong interaction.  The problem is characterized by 
the $\Delta T=1/2$ selction rule, best known as the fact that the isospin 
$T=0$ amplitude of the $K\rightarrow \pi\pi$ decay is about 22 times 
larger than the $T=2$ amplitude.  In spite of many efforts this 
enhancement of the $\Delta T=1/2$ channel over the $\Delta T=3/2$ channel 
has not been explained in a satisfactory way.  A part of the answer comes 
from perturbative gluons which are created if one evolves the simple 
$W$-exchange-vertex from a scale of 80 GeV down to 
1 GeV~\cite{GaillardLee,AM,Witten,VZS,SVZ,GW,GP,BW,Burasetal}.  Another 
part of the answer is supposed to arise from the structure of the light 
hadrons, whose description at scales around $1$ GeV requires a
nonperturbative QCD-method.  

In the low energy regime one way to deal 
with nonperturbative effects is to utilize the large $N_c$ expansion 
with $\alpha_s N_c$ fixed ($N_c$ being the number of colors and 
$\alpha_s$ the running coupling constant of QCD).  In the limit of 
large $N_c$ QCD can be treated as a weakly coupled meson field theory 
and indeed many experimental consequences have been explained in this 
way~\cite{tHooft,Witten2}.  The large $N_c$ limit of QCD was also 
employed~\cite{FYSY,NV,TT} in order to understand the $\Delta T=1/2$ problem 
in the $K\rightarrow \pi\pi$ decay. However, in contrast to the 
sector of the pure strong interaction, the large $N_c$ limit in its strict 
form (only leading order in $N_c$) does not seem to be sufficient 
to describe the weak non-leptonic decays~\cite{CFG} because it enhances 
the $\Delta T=3/2$ channel while suppressing the $\Delta T=1/2$
one making the problem even more difficult.  Hence for these processes one 
is bound to go beyond leading order in the large $N_c$ expansion.  

At low energies chiral perturbation theory ($\chi$PT)~\cite{GasserLeutwyler}
is known as a proper effective field theory of QCD in the mesonic 
sector.  Based on its success in describing strong interactions $\chi$PT 
was also applied to nonleptonic processes of light 
mesons~\cite{Kamboretal,Esposito,EKW}.  However in this case there are not 
enough experimental data available to determine the many low energy 
constants (LECs) of the the effective weak chiral Lagrangian to order 
${\cal O}(p^4)$.  Hence, in order to proceed without experiments, 
one is advised to determine the LECs of the weak chiral Lagrangian by 
using effective QCD-inspired models.

   In the present paper we are going to investigate how far the chiral 
quark model ($\chi$QM) furnishes a reasonable framework to determine the 
LECs of the weak chiral Lagrangian.  We are motivated to this study 
by success of the $\chi$QM to determine the LECs of the strong chiral 
Lagrangian.  The $\chi$QM is characterized by the Euclidean partition 
function~\cite{DPP}
\begin{equation}
{\cal Z}\;=\; \int {\cal D} \psi {\cal D} \psi^\dagger 
{\cal D} \pi
\exp \left[\int d^4 x \psi^{\dagger \alpha}_{f}
\left(i\rlap{/}{\partial} + 
iM e^{i\gamma_5\lambda^a \pi^a} \right)_{fg}
\psi^{\alpha}_{g} \right],
\label{Eq:part}
\end{equation}
where $\alpha$ is the color index, $\alpha = 1,\cdots, N_c$ and 
$f$ and $g$ are flavor indices.  The $M$ serves as the coupling parameter 
between the constituent quark fields $\psi$ and the Goldstone boson 
field $\pi^a$ and it can be identified with the constituent quark mass.  In 
this work we want to construct systematically the effective weak chiral 
Lagrangian without external fields to order ${\cal O}(p^2)$ and 
${\cal O} (p^4)$ for $\Delta S =1$ and $\Delta S = 2$.  We will show that 
the $\chi$QM provides the most general structure of the Lagrangian, 
known from the work of Refs.~\cite{Kamboretal,Esposito,EKW}, 
and unique descriptions of the LECs in leading and subleading order of 
the large $N_c$ expansion.  The calculations start from the effective weak 
Hamiltonians for $\Delta S=1,2$ of Refs.~\cite{Burasetal,BJW,HN1,HN2}.
 
    The effective weak Lagrangian for $\Delta S = 1$ 
to order ${\cal O}(p^2)$ in leading and subleading order in 
$N_c$ has been given already by Antonelli {\em et al.}
\cite{Antonellietal}.  Bertolini {\em et al.}~\cite{Bertolinietal}
extended the former calculation to ${\cal O}(p^4)$ in the study of 
$\epsilon'/\epsilon$ and $\hat{B}_K$ with the $\Delta S = 1$ Lagrangian, 
which implies that parts of the ${\cal O} (p^4)$ effective chiral weak 
Lagrangian that are necessary for the description of the 
$K\rightarrow \pi\pi$ decays are obtained.  They used for this the small 
field expansion in leading and subleading order in the $N_c$ 
expansion.  In the present paper we prefer the derivative expansion, 
since we want to evaluate the full 
effective chiral weak Lagrangian to order ${\cal O}(p^4)$ including 
the corresponding LECs in a way that it can be used directly in 
$\chi$PT.  This means we have to use an expansion of the $\chi$QM 
which is consistent with the chiral expansion in $\chi$PT.  In 
the case of the strong interaction it is known that the small 
field 
expansion fulfills this criterion only in the leading order in the large $N_c$ 
expansion.  However, for the weak chiral Lagrangian the subleading order 
in $N_c$ is necessary and hence the derivative expansion seems to us more 
appropriate than the small field expansion~\footnote{The LECs
$L_1$ and $L_2/2$ of the Gasser-Leutwyler 
Lagrangian~\cite{GasserLeutwyler} in the strong interaction 
are the same in the large $N_c$ limit, which makes 
the small field expansion in the chiral quark model yield the same Lagrangian 
as in the derivative 
expansion.  However, when one considers higher order 
corrections $L_1$ and $L_2/2$ have to deviate from each other, as the 
experimental extraction of those values implies, a feature, which is 
only brought out by the derivative expansion.}.  

   The outline of the present paper is as follows: In section 2 we sketch 
the characteristics of the chiral quark model and briefly show how to use 
the derivative expansion.  In section 3 we review the effective weak chiral 
action at a scale of 1 GeV and discuss some of its properties relevant for 
the following.  Section 4 is devoted to the derivation of the effective weak 
chiral Lagrangian to order ${\cal O} (p^2)$.  We examine the dependence of 
the LECs on the constituent quark mass, the quark condensate and the quark
axial-vector constant.  The full effective weak chiral 
Lagrangian for $\Delta S=1$ and $\Delta S=2$ to order ${\cal O}(p^4)$ 
in leading and next-to-leading order in $N_c$, as it results from the 
derivative expansion, is presented in Section 5.  The conclusions are given 
in Section 6.
\section{Chiral quark model}
The characteristic of the chiral quark model is represented by
the effective chiral action in Euclidean
space given by the functional integral over quark fields~\cite{DPP}:
\begin{equation}
{\cal N} \;=\; \exp{\left(-S_{\rm eff}\right)}\;=\; 
\int {\cal D} \psi {\cal D} \psi^\dagger 
\exp \left[\int d^4 x \psi^{\dagger \alpha}_{f}
\left(i\rlap{/}{\partial}  + iM U^{\gamma_5} \right)_{fg}
\psi^{\alpha}_{g} \right],
\label{Eq:effect}
\end{equation}
where $\alpha$ is the color index, $\alpha = 1,\cdots, N_c$ and 
$f$ and $g$ are flavor indices.
$M$ is the constituent quark mass, which is in fact momentum-dependent.
However, we regard it as a free parameter for convenience 
and introduce a cut-off parameter to tame the divergence appearing in 
the quark loop.  
It is fixed by producing the pion decay constant. 
$U^{\gamma_5}$ denotes the Goldstone field 
\begin{equation}
U^{\gamma_5}\;=\;\exp\left(i\pi^a\lambda^a\gamma_5\right)
\;=\;  U\frac{1+\gamma_5}{2} + U^\dagger \frac{1-\gamma_5}{2}
\end{equation} 
with
\begin{equation}
U\;=\; \exp{\left(i\pi^a \lambda^a\right)}.
\end{equation}
The $\pi$ stands for the meson octet fields
\begin{equation}
\pi\;=\;\pi^a \lambda^a = \frac{1}{\sqrt{2}}\left(
\begin{array}{ccc}
\frac{1}{\sqrt{2}}\pi^0 + \frac{1}{\sqrt{6}} \eta & \pi^+ & K^+ \\
\pi^- & -\frac{1}{\sqrt{2}}\pi^0 + \frac{1}{\sqrt{6}} \eta & K^0 \\
K^-&\bar{K}^0& -\frac{1}{\sqrt{2}}\pi^0 + \frac{1}{\sqrt{6}} \eta 
\end{array}\right).
\end{equation}

Integrating over the quark fields in Eq.(\ref{Eq:effect}),
we obtain the following expression for the effective chiral action
\begin{equation}
S_{\rm eff} \left[ U \right] \;=\; -N_c {\rm Tr} \ln  D
\label{Eq:echa}
\end{equation}
where ${\rm Tr}$ designates the functional trace as well as flavor and spin
ones.  The $D$ denotes the Dirac operator
\begin{equation}
D \;=\; i\rlap{/}{\partial} 
+ iM U^{\gamma_5} .
\label{Eq:Dirac}
\end{equation}      
Since Eq.(\ref{Eq:echa}) is non-Hermitian, one can separate the
effective action into the real part and the imaginary one.  The real
part can be written as
\begin{equation}
{\rm Re} S_{\rm eff} \left[U \right] \;=\; -\frac12 {\rm Tr} \ln  
\left(\frac{D^\dagger D}{D^\dagger_{0} D_0}\right),
\end{equation}
where
\begin{eqnarray}
D^\dagger D &=& -\partial^2 + M^2 - M\left(\rlap{/}{\partial} U^{\gamma_5}
\right),
\nonumber \\
D^\dagger_{0} D_{0} &=& -\partial^2 + M^2 .
\end{eqnarray}
It is already well known how to treat the effective action in order to
obtain the effective chiral Lagrangian~\cite{DE,AF,chan,dsw}.  The
real part of the effective action may be expanded with respect to the
derivatives of the meson field~\cite{DPP}:
\begin{eqnarray}
{\rm Re} S_{\rm eff} &=& -\frac{N_c}{2} {\rm Tr} \ln 
\left( 1 - \frac{M(\rlap{/}{\partial}U^{\gamma_5})}
{D^\dagger_{0} D_0}\right)\nonumber \\
&=&-\frac{N_c}{2} {\rm tr} \int d^4 x \left\langle x \left |  
\ln \left(1 - \frac{M(\rlap{/}{\partial}U^{\gamma_5})}
{D^\dagger_{0} D_0}\right)
\right| x\right\rangle \nonumber \\
&=& -\frac{N_c}{2} \int d^4 x \int \frac{d^4 k}{(2\pi)^4} 
{\rm tr} \ln \left( 1 - \frac{M(\rlap{/}{\partial}U^{\gamma_5})}
{k^2 +M^2 - (2ik\cdot \partial + \partial^2)}\right)\cdot 1
\end{eqnarray}
The nonvanishing leading term in the expansion is just the kinetic 
Lagrangian of the strong interaction:
\begin{equation}
{\rm Re} S ^{(2)}_{\rm eff}[U] \;=\; -  
\int d^4 x {\cal L}^{(2)}_{\rm \Delta S = 0},
\end{equation}
where
\begin{equation}
{\cal L}^{(2)}_{\rm \Delta S = 0} = - \frac{f^2_{\pi}}{4} 
\left\langle L_\mu L_\mu \right\rangle
\label{Eq:strong}
\end{equation}
The $L_\mu$ ($R_\mu$) are the Noether currents of 
${\rm SU(3)}_{L} \times {\rm SU(3)}_{R}$ chiral symmetry:
\begin{equation}
L_\mu \;=\; iU^\dagger \partial_\mu U,\;\;\;
R_\mu \;=\; iU \partial_\mu U^\dagger.
\end{equation}
The symbol $\langle \rangle$ stands for the flavor trace.
The $f_{\pi}$ denote the pion decay constant which is 
related to the following quark loop integral:
\begin{equation}
f^2_{\pi} \;=\; 4 N_c \int \frac{d^4 k}{(2\pi)^4}
\frac{M^2}{(k^2 + M^2)^2}.
\label{Eq:fq}
\end{equation}
Since the quark loop integral is divergent, which is due to
the fact that we regard the $M$ as a constant, we need to introduce the
cut-off parameter $\Lambda$ via regularization.  It is fixed by producing
the experimental value of $f_\pi = 93$ MeV.

Similarly, we can move up to higher orders in the derivative 
expansion.  The real part of the effective chiral action in the next-to-leading 
order ${\cal O}(p^4)$ is given by 
\begin{equation}
{\rm Re} S ^{(4)}_{\rm eff} \;=\; -\int d^4 x {\cal L}^{(4)}_{\rm \Delta S = 0},
\end{equation}
so that the strong effective chiral Lagrangian to order ${\cal O}(p^4)$
can be written as~\cite{DPP}
\begin{equation}
{\cal L}^{(4)}_{\rm \Delta S = 0} = \frac{N_c}{192\pi^2}
\int d^4x \left[2\langle (\partial_\mu L_\mu)^2\rangle +
\langle L_\mu L_\nu L_\mu L_\nu\rangle\right].
\end{equation}
Those effective Lagrangians in higher orders 
were extensively studied~\cite{DE,AF,chan,dsw,Zuk,ERT,Bijnens}.  Ref.~\cite{ERT} 
investigated also the low energy constants of the 
effective ${\cal O} (p^4)$ Lagrangian in relation to chiral perturbation 
theory.

The imaginary part of the effective chiral action is pertinent to the
Wess-Zumino-Witten(WZW) action~\cite{WZ,Witten3} with the correct
coefficient, which arises from the derivative expansion of the imaginary part 
to order ${\cal O}(p^5)$ (see Ref.~\cite{DPP} for details.).      
    
\section{Effective weak chiral action}
The effective chiral action in Eq.(\ref{Eq:part}) with the 
weak $\Delta S = 1$ or $\Delta S = 1$ effective Hamiltonian can be 
written as follows: \begin{equation}
\exp{\left(- S^{\Delta S = 1,2}_{\rm eff}\right)} \;=\; 
\int {\cal D} \psi {\cal D} \psi^\dagger 
\exp \left[\int d^4 x \left(\psi^{\dagger} D
\psi - {\cal H}^{\Delta S = 1,2}_{\rm eff}\right)
\right],
\label{Eq:partw}
\end{equation}
Here the effective weak quark Hamiltonian 
${\cal H}^{\Delta S = 1}_{\rm eff}$ consists of ten four-quark operators 
among which only seven operators are independent
\begin{equation}
{\cal H}^{\Delta S = 1}_{\rm eff}
 \;=\; -\frac{G_F}{\sqrt{2}} V_{ud} V^*_{us}
\sum_i c_i (\mu) {\cal Q}_i (\mu) + {\rm h.c.} .
\end{equation}
The $G_F$ is the well-known Fermi constant and $V_{ij}$ denote
the Cabibbo-Kobayashi-Maskawa(CKM) matrix elements.
The $\tau$ is their ratio
given by $\tau = - V_{td} V^*_{ts}/V_{ud} V^*_{us}$.
The $c_i(\mu)$ consist of the Wilson coefficients:
$c_i (\mu) = z_i (\mu) + \tau y_i (\mu)$.  
The functions $z_i (\mu)$ and $y_i (\mu)$ are the scale-dependent
Wilson coefficients given at the scale of the $\mu$. The 
$z_i (\mu)$ represent the $CP$-conserving part,
while $y_i (\mu)$ stand for the $CP$-violating one.  
The four-quark operators ${\cal Q}_i$ contain the dynamic information
of the weak transitions, being constructed by
integrating out the vector bosons $W^{\pm}$ and $Z$ and
heavy quarks $t$, $b$ and $c$.  
The four-quark operators~\cite{Burasetal} are given by 
\begin{eqnarray}
{\cal Q}_1  & = & 4 \left( {s}^\dagger_{\alpha} 
\gamma_{\mu} P_L u_{\beta} \right)
     \left( {u}^\dagger_{\beta} \gamma_{\mu} P_L  d_{\alpha} \right)
,\label{Eq:qqi} \\
{\cal Q}_2  & = & 4 \left( {s}^\dagger_{\alpha} 
\gamma_{\mu} P_L u_{\alpha} \right)
    \left( {u}^\dagger_{\beta} \gamma_{\mu} P_L d_{\beta}  \right)
, \\
{\cal Q}_3 & = & 4 \left( {s}^\dagger_{\alpha} \gamma_{\mu} 
P_L d_{\alpha} \right)
  \sum_{q=u,d,s} \left( {q}^\dagger_{\beta} \gamma_{\mu} P_L  q_{\beta} \right)
, \\
{\cal Q}_4 & = & 4 \left( {s}^\dagger_{\alpha} 
\gamma_{\mu} P_L d_{\beta} \right)
  \sum_{q=u,d,s} \left( {q}^\dagger_{\beta} \gamma_{\mu} P_L q_{\alpha} \right)
, \\
{\cal Q}_5 & = & 4 \left( {s}^\dagger_{\alpha} 
\gamma_{\mu} P_L d_{\alpha} \right)
  \sum_{q=u,d,s} \left( {q}^\dagger_{\beta} \gamma_{\mu} P_R q_{\beta} \right)
, \\
{\cal Q}_6  & = & 4 \left( {s}^\dagger_{\alpha} 
\gamma_{\mu} P_L d_{\beta} \right)
  \sum_{q=u,d,s} \left( {q}^\dagger_{\beta} \gamma_{\mu} P_R q_{\alpha} \right)
, \\
{\cal Q}_7 & = & 6 \left( {s}^\dagger_{\alpha} 
\gamma_{\mu} P_L d_{\alpha} \right)
 \sum_{q=u,d,s} 
\left( {q}^\dagger_{\beta} \hat{Q} \gamma_{\mu} P_R q_{\beta} \right)
, \\
{\cal Q}_8 & = & 6 \left( {s}^\dagger_{\alpha} 
\gamma_{\mu} P_L d_{\beta} \right)
 \sum_{q=u,d,s} 
\left( {q}^\dagger_{\beta} \hat{Q} \gamma_{\mu} P_R q_{\alpha} \right)
, \\
{\cal Q}_9 & = & 6 \left( {s}^\dagger_{\alpha} 
\gamma_{\mu} P_L d_{\alpha} \right)
  \sum_{q=u,d,s} 
\left( {q}^\dagger_{\beta} \hat{Q} \gamma_{\mu} P_L q_{\beta}  \right)
, \\
{\cal Q}_{10} & = & 6 \left( {s}^\dagger_{\alpha} 
\gamma_{\mu} P_L d_{\beta} \right)
  \sum_{q=u,d,s} 
\left( {q}^\dagger_{\beta} \hat{Q} \gamma_{\mu} P_L q_{\alpha}  \right) ,
\label{Eq:qq}
\end{eqnarray}
where $P_{L,R} = \frac{1}{2} \left( 1 \pm \gamma_5 \right)$ are the chiral 
projection operators and $\hat{Q}=\frac13 {\rm diag}(2,-1,-1)$ denote the 
quark charge matrix. The ${\cal Q}_1$ and ${\cal Q}_2$ come from the 
current-current diagrams, while ${\cal Q}_3$ to ${\cal Q}_6$~\cite{SVZ,GW,GP} 
and ${\cal Q}_7$ to ${\cal Q}_{10}$~\cite{BW}
are induced by QCD penguin and electroweak penguin
diagrams, respectively.  Note that only seven operators in 
Eqs.(\ref{Eq:qqi}-\ref{Eq:qq})
are independent.  For example, we can express 
${\cal Q}_4$, ${\cal Q}_9$, and ${\cal Q}_{10}$ as follows:
\begin{equation}
{\cal Q}_4 = - {\cal Q}_1 + {\cal Q}_2 + {\cal Q}_3,\;\;\;
{\cal Q}_9 = \frac12 \left(3{\cal Q}_1 - {\cal Q}_3\right),\;\;\;
{\cal Q}_{10} = {\cal Q}_2 + 
\frac12 \left({\cal Q}_1 - {\cal Q}_3\right).
\end{equation}
Under the chiral transformation ${\rm SU}(3)_L \times {\rm SU(3)}_R$
the four-quark operators ${\cal Q}_{3,4,5,6}$ transform like
$(\underline{8}_L, \underline{1}_R)$.
  The ${\cal Q}_{1,2,9,10}$ transform 
like the combination of $(\underline{8}_L, \underline{1}_R)$ 
and $(\underline{27}_L, \underline{1}_R)$, 
while the ${\cal Q}_{7,8}$ transform like 
$(\underline{8}_L, \underline{8}_R)$. 
 The $\Delta S = 2$ effective 
weak Hamiltonian is expressed as~\cite{GilmanWise2,BJW,HN1,HN2}
\begin{equation}
{\cal H}^{\Delta S = 2}_{\rm eff} \;=\;-\frac{G^2_{F} M^2_{W}}{16\pi^2}
{\cal F} \left(\lambda_c,\lambda_t,m^2_{c},m^2_{t},M^2_{W}\right)b(\mu) 
{\cal Q}_{\Delta S = 2} (\mu) + \mbox{h.c.}
\end{equation}
with
\begin{equation}
{\cal F} \;=\;\lambda^2_{c} \eta_1  S \left(\frac{m^2_{c}}{M^2_{W}}\right) +
\lambda^2_{t} \eta_2  S \left(\frac{m^2_{t}}{M^2_{W}}\right)
+ 2 \lambda_{c}\lambda_{t} \eta_3  S 
\left(\frac{m^2_{c}}{M^2_{W}},\frac{m^2_{t}}{M^2_{W}}\right)
\end{equation}
and the parameters $\lambda_q=V_{qd}V^{*}_{qs}$ denote the 
pertinent relations of the CKM matrix elements with $q=u,c,t$.  The functions
$S_{i}$ are the Inami-Lim functions~\cite{GilmanWise2,InamiLim,BBH},
being obtained by integrating over electroweak loops
and describing the $|\Delta S| = 2$ transition amplitude in the absence of 
strong interactions.  The $b(\mu)$ is again the corresponding Wilson 
coefficient.  The coefficients $\eta_i$ represent the short-distance QCD 
corrections split off from the $b(\mu)$~\cite{HN2}.  The four-quark operator 
${\cal Q}_{\Delta S = 2}$ is written as
\begin{equation}
{\cal Q}_{\Delta S = 2} \;=\; 4 \left(s^{\dagger}_\alpha
\gamma_\mu P_L d_\alpha\right)\left(s^{\dagger}_\beta \gamma_\mu P_L
d_\beta\right). 
\end{equation}

Since the Fermi constant $G_F$ is very small, 
one can expand Eq.(\ref{Eq:partw}) in powers of the $G_F$ and keep the
lowest order only.  Then we can obtain the effective weak chiral 
Lagrangian
\begin{equation}
{\cal L}^{\Delta S = 1,2}_{\rm eff} \;=\; -\frac{1}{\cal N}
\int {\cal D} \psi {\cal D} 
\psi^\dagger {\cal H}^{\Delta S = 1,2}_{\rm eff} \exp  
\left[\int d^4 x \psi^\dagger D \psi\right].
\label{Eq:part1}
\end{equation}

If you write a generic operator for the four-quark operator
${\cal Q}_i$ for a given $i$ in Euclidean space such as
\begin{equation}
{\cal Q}_i(x) \;=\; \psi^\dagger (x) \gamma_\mu {P}_{R,L} \Lambda_1
\psi(x) \psi^\dagger (x) \gamma_\mu {P}_{R,L} \Lambda_2 \psi(x),
\end{equation}  
where ${\Lambda}_{1,2}$ denote the flavor 
spin operators, then we can calculate the vacuum expectation value (VEV)
of ${\cal Q}_i(x)$ as follows:
\begin{eqnarray}
\langle {\cal Q}_i \rangle &=& \frac{1}{\cal N}\int {\cal D} \psi {\cal D} 
\psi^\dagger {\cal Q}_i(x) \exp \left[\int d^4 z \psi^\dagger D \psi\right]
\nonumber \\
&=& \int d^4y \int \frac{d^4 k}{(2\pi)^4} 
e^{ik(x-y)} 
\frac{\delta}{\delta J^{(1)}_{\mu} (x)} \frac{\delta}{\delta J^{(2)}_\mu (y)}
\nonumber \\
&& \times 
\exp \left[\int d^4z \left \langle z \left| {\rm tr} \ln \tilde{D}
(J_1(z),J_2(z))\right| z\right\rangle \right]_{J_{1}=J_{2}=0} \nonumber \\
&=& L^{(1)}_i \;+\; L^{(2)}_i.
\end{eqnarray}
Here, $\tilde{D}$ is 
\begin{equation}
\tilde{D}(J_1(z),J_2(z)) \;=\; D + J^{(1)}_{\alpha}(z) \gamma_\alpha
{P}_{R,L} \Lambda_1
+ J^{(2)}_{\beta} (z) \gamma_\beta {P}_{R,L} \Lambda_2.
\end{equation}
The $L^{(1)}_i$ and $L^{(2)}_i$ are given by
\begin{eqnarray}
L^{(1)}_i &= & -N^{2}_c  {\rm tr} \left[\left\langle x\left| 
\frac{1}{D} \gamma_\mu 
{P}_{R,L} \Lambda_1 \right |x \right \rangle \left\langle x \left |
\frac{1}{D}\gamma_\mu 
{P}_{R,L} \Lambda_2 \right |x \right \rangle \right]_i  
+ {\cal O}\left( N_c \right) \nonumber \\
&=& -N_c^2 {\rm tr} \left[ {(A_1)}_\mu {(A_2)}_\mu
\right]_i 
+ {\cal O}\left( N_c \right) \hspace{19mm} i=1,4,6,8,10 \; , 
\label{Eq:wea1}\\
L^{(2)}_i &= & N^{2}_c {\rm tr} \left[ \left \langle x\left | 
\frac{1}{D}\gamma_\mu 
{P}_{R,L} \Lambda_1 \right | x \right 
\rangle \right]{\rm tr} \left[\left \langle x \left |\frac{1}{D}
\gamma_\mu 
{P}_{R,L} \Lambda_2 \right | x \right \rangle \right]_i
+ {\cal O}\left( N_c \right) \nonumber \\
&=& N_c^2 {\rm tr} \left[ {(A_1)}_{\mu} \right]
 {\rm tr} \left[  {(A_2)}_{\mu} \right]_i 
+ {\cal O}\left( N_c \right) \hspace{10mm} 
i=2,3,5,7,9 \;,
\label{Eq:wea2}
\end{eqnarray}
where $\Lambda_{1,2}$ are the corresponding flavor matrices.
The operators ${(A_{1,2})}_\mu$ can be written as
\begin{equation}
{(A_{1,2})}_{\mu} \;=\; \int {d^4k \over (2 \pi)^4} \;
\left[
\frac{ i\!\not\!\partial + \not\!k 
- i M U^{-\gamma_5}}{k^2 + M^2 - \partial^2 + 2 i k\cdot \partial -
M (\!\not\!\partial U^{\gamma_5}) }\right] \gamma_{\mu}
P_{L,R} \Lambda_{1,2} .
\label{Eq:A12}
\end{equation}
Assuming that the pion field changes adiabatically, we are able to 
expand the denominator in Eq.(\ref{Eq:A12}) in powers of 
$\partial^2 - 2 i k \cdot \partial + M (\!\not\!\partial U^{\gamma_5})$
and obtain the following expression:
\begin{eqnarray}
{(A_{1,2})}_{\mu}
&=& \int {d^4k \over (2 \pi)^4} \;
{1 \over k^2 + M^2} \sum_{n=0}^{\infty}
\left[ { \partial^2 - 2 i k \cdot \partial + 
M (\!\not\!\partial U^{\gamma_5})
\over k^2 + M^2} \right]^n \nonumber \\ &\times&
\left( i\!\not\!\partial + \not\!k - i M U^{-\gamma_5} \right)
 \gamma_{\mu} P_{L,R} \Lambda_{1,2} .
\label{Eq:grad}
\end{eqnarray}
With the expansion given in Eq.(\ref{Eq:grad}) we can systematically
evaluate effective weak chiral Lagrangian to order ${\cal O} (p^4)$:
\begin{equation}
{\cal L}^{\Delta S = 1,2}_{\rm eff} \;=\; 
{\cal L}^{\Delta S = 1,2}_{\rm eff} ({\cal O}(p^2))
\;+\;  {\cal L}^{\Delta S = 1,2}_{\rm eff} ({\cal O}(p^4)). 
\label{Eq:laq}
\end{equation}
We first evaluate the effective weak chiral Lagrangian in the lowest 
order.    
\section{Lowest order $p^2$ and low energy constants}
\subsection{Leading order in the $1/N_c$ expansion}
The derivation of the ${\cal L}^{\Delta S = 1,2}_{\rm eff}({\cal O}(p^2))$ 
is straightforward.  At lowest leading order in the derivative expansion, 
{\em i.e.} ${\cal O} (p^2)$ order, we obtain the following results
with ${\cal O} (N^{2}_c)$ considered:
\begin{eqnarray}
\langle {\cal Q}_1+{\cal Q}^{\dagger}_1 \rangle_{{\cal O} (p^2)} 
& = & f^4_{\pi} \left(-\frac25 \langle \lambda_6 L_\mu L_\mu 
\rangle + \frac13 t_{ij;kl} \langle \lambda_{ij} L_\mu \rangle 
\langle \lambda_{kl} L_\mu \rangle\right),  \label{Eq:qa}\\
\langle {\cal Q}_2+{\cal Q}^{\dagger}_2\rangle_{{\cal O} (p^2)} 
& = & f^4_{\pi} \left(\frac35 \langle \lambda_6 L_\mu L_\mu 
\rangle + \frac13 t_{ij;kl} \langle \lambda_{ij} L_\mu \rangle 
\langle \lambda_{kl} L_\mu \rangle\right),\\
\langle {\cal Q}_3+{\cal Q}^{\dagger}_3\rangle_{{\cal O} (p^2)}  & = & 0,\\
\langle {\cal Q}_4+{\cal Q}^{\dagger}_4\rangle_{{\cal O} (p^2)}  
& = & f^4_{\pi} 
\langle \lambda_6  L_\mu  L_\mu \rangle,  \\
\langle {\cal Q}_5+{\cal Q}^{\dagger}_5\rangle_{{\cal O} (p^2)} & = & 0,\\
\langle {\cal Q}_6+{\cal Q}^{\dagger}_6\rangle_{{\cal O} (p^2)} & = & 
\left(\frac{\langle \bar{q} q \rangle f^2_{\pi}}{M}
-\frac{\langle \bar{q} q \rangle N_c M}{8 \pi^2}\right) 
f^4_{\pi} 
\langle \lambda_6  L_\mu  L_\mu \rangle,  \\
{\langle {\cal Q}_7+{\cal Q}^{\dagger}_7 \rangle}_{{\cal O}(p^2)} &=& 
{3 \over 2} f_\pi^4 
\langle L_\mu \lambda_6 \rangle \langle R_\mu \hat{Q} \rangle,\\
{\langle {\cal Q}_8+{\cal Q}^{\dagger}_8 \rangle}_{{\cal O}(p^2)} &=&  - 
\left( { N_c \langle \bar{q}q\rangle M \over 16 \pi^2 }
+ { f_\pi^2 \langle \bar{q}q\rangle \over 2 M } \right) 
\Big[ 
\langle U \lambda_6 \left( \partial^2 U^\dagger \right) \hat{Q} \rangle
+ \langle \left( \partial^2 U \right) \lambda_6 U^\dagger 
\hat{Q} \rangle \Big]\nonumber  \\
&&\hspace{-1.5cm} - { N_c \langle \bar{q}q\rangle M \over 8 \pi^2 }
\Big[ 
\langle U \lambda_6 \left( \partial_\mu U^\dagger \right)
\left( \partial_\mu U \right) U^\dagger  \hat{Q} \rangle
+ \langle  \left( \partial_\mu U \right)
\left( \partial_\mu U^\dagger \right) U \lambda_6 U^\dagger 
\hat{Q} \rangle \Big], \\
\langle {\cal Q}_9+{\cal Q}^{\dagger}_9\rangle_{{\cal O} (p^2)}
& = & f^4_{\pi} 
\left(-\frac35 \langle \lambda_6 L_\mu L_\mu 
+ \frac12 t_{ij;kl} \langle \lambda_{ij} L_\mu \rangle 
\langle \lambda_{kl} L_\mu \rangle\right),\\
\langle {\cal Q}_{10}+{\cal Q}^{\dagger}_{10}\rangle_{{\cal O} (p^2)} & = & 
 f^4_{\pi} 
\left(\frac25 \langle \lambda_6 L_\mu L_\mu 
+ \frac12 t_{ij;kl} \langle \lambda_{ij} L_\mu \rangle 
\langle \lambda_{kl} L_\mu \rangle\right), \\
\langle {\cal Q}_{\Delta S = 2} + {\cal Q}^{\dagger}_ {\Delta S = 2}
\rangle_{{\cal O} (p^2)} &=& f^4_{\pi}
\langle \lambda_6 L_\mu \lambda_6 L_\mu \rangle .
\label{Eq:qz}
\end{eqnarray}
The eikosiheptaplet projection operators $t_{ij;kl}$
are defined by 
\begin{equation}
t_{ij;kl} \;=\; \frac 15 t^{T=1/2}_{ij;kl} + t^{T=3/2}_{ij;kl},
\end{equation}
where
\begin{eqnarray}
t^{T=1/2}_{13;21} &=& t^{T=1/2}_{31;12}\;=\;t^{T=1/2}_{21;13}
\;=\;t^{T=1/2}_{12;31}\;=\;\frac12,\nonumber \\
t^{T=1/2}_{23;11} &=& t^{T=1/2}_{32;11}\;=\;t^{T=1/2}_{11;23}
\;=\;t^{T=1/2}_{11;32}\;=\;\frac12,\nonumber \\
t^{T=1/2}_{23;22} &=& t^{T=1/2}_{32;22}\;=\;t^{T=1/2}_{22;23}
\;=\;t^{T=1/2}_{22;32}\;=\;1,\nonumber \\
t^{T=1/2}_{23;33} &=& t^{T=1/2}_{32;33}\;=\;t^{T=1/2}_{33;23}
\;=\;t^{T=1/2}_{33;32}\;=\;-\frac32,\nonumber \\
t^{T=3/2}_{13;21} &=& t^{T=3/2}_{31;12}\;=\;t^{T=3/2}_{21;13}
\;=\;t^{T=3/2}_{12;31}\;=\;\frac12,\nonumber \\
t^{T=3/2}_{13;21} &=& t^{T=3/2}_{31;12}\;=\;t^{T=3/2}_{21;13}
\;=\;t^{T=3/2}_{12;31}\;=\;\frac12,\nonumber \\
t^{T=3/2}_{13;21} &=& t^{T=3/2}_{31;12}\;=\;t^{T=3/2}_{21;13}
\;=\;t^{T=3/2}_{12;31}\;=\;-\frac12,\nonumber \\
t^{T=1/2}_{ij;kl} &=& t^{T=3/2}_{ij;kl} \;=\; 0,\;\;\;\mbox{for
the other } i,j,k,l,
\end{eqnarray}
and
\begin{equation}
(\lambda_{ij})_{ab} \;=\; \delta_{ia} \delta_{ib}.
\end{equation}
The coefficients appearing in front of the integrals consist
of the pion decay constant $f_\pi$ (see Eq.(\ref{Eq:fq})), 
quark condensate\footnote{
In our calculation the quark condensate is defined 
as $\langle \bar{q} q\rangle=\langle \bar{u} u + \bar{d} d \rangle$,
 since it plays the role 
of a numerical parameter in our calculation we 
do not distinguish between the quark condensate in Euclidean 
and Minkowski space.} 
$\langle \bar{q} q\rangle$, and constituent quark mass $M$.  The
quark condensate is related to the following quadratically-divergent
integral: 
\begin{equation}
\langle\bar{q} q\rangle  \;=\; 8 N_c \int \frac{d^4 k}{(2\pi)^4}
\frac{M}{k^2 + M^2}.
\end{equation}
Hence, we find that those coefficients have ${\cal O} (N_c^{2})$ order
in the $N_c$ counting.  Note that the VEV of the operators
$\langle {\cal Q}_{3}\rangle$ and $\langle {\cal Q}_{5}\rangle$
vanish at leading order in $N_c$,    
which implies that in the leading order of the large $N_c$ expansion 
$\langle {\cal Q}_{3}\rangle$ and $\langle {\cal Q}_{5}\rangle$ 
do not contribute to the effective weak chiral Lagrangian in the 
order ${\cal O} (p^2)$.  
  
It is interesting to compare our results with those of 
Ref.~\cite{Antonellietal}.  We find some differences in 
$\langle {\cal Q}_{6}\rangle$ and $\langle {\cal Q}_{8}\rangle$ which, 
however, disappear when we apply for the quark condensate the same 
regularization scheme as used in  Ref.~\cite{Antonellietal}. 
Since Ref.~\cite{Antonellietal} employs the expansion of the 
weak meson field in which the $U$ field is expanded in powers of
the $\pi$ field, one is not able to obtain the full effective chiral 
Lagrangian to order ${\cal O}(p^4)$ consistently with the chiral expansion, 
if one goes beyond the leading 
order in the large $N_c$ expansion.  With the derivative expansion, 
we can derive the effective weak chiral Lagrangian in the next-to-leading 
order ${\cal O} (p^4)$.  In fact, the VEV of the operator ${\cal Q}_8$ has 
the zeroth order contribution:
\begin{equation}
\langle {\cal Q}_{8}+{\cal Q}^{\dagger}_{8}\rangle_{{\cal O}(p^0)} \;=\;
\frac34 {\langle \bar{q} q \rangle}^2 \langle 
\lambda_6 U^\dagger \hat{Q} U\rangle .
\end{equation}
It transforms under ${\rm SU}(3)_{L} \times {\rm SU}(3)_{R}$ as
$(\underline{8}_{L},\, \underline{8}_{R})$.  

The effective weak chiral Lagrangian describing the $\Delta S = 1$
nonleptonic decays of kaons was first introduced by Cronin~\cite{Cronin} 
(presented in Minkowski space):
\begin{eqnarray}
{\cal L}^{\Delta S = 1,{\cal O} (p^2)}_{\rm eff}&=& 
-\frac{G_F}{\sqrt{2}} V_{ud} V^*_{us} f^4_{\pi} 
\left[g_{\underline{8}} \left\langle\lambda_{23} L_\mu L^\mu \right\rangle
\right.\nonumber \\ && \;+\; \left.g_{\underline{27}} \left(\frac23
\left\langle \lambda_{12} L_\mu\right\rangle
\left\langle \lambda_{31} L^\mu \right\rangle  + 
\left\langle \lambda_{32} L_\mu\right\rangle
\left\langle \lambda_{11} L^\mu \right\rangle \right) \right]
\;+\; {\rm h.c.}  \nonumber \\ 
&=& {\cal L}^{(1/2)}_{\underline{8}} \;+\; \frac19 
{\cal L}^{(1/2)}_{\underline{27}} \;+\;  
\frac59 {\cal L}^{(3/2)}_{\underline{27}},
\label{Eq:Lcpt2}
\end{eqnarray}
where
\begin{eqnarray}
{\cal L}^{(1/2)}_{\underline{8}} &=& 
-\frac{G_F}{\sqrt{2}} V_{ud} V^*_{us}f^4_{\pi} 
g_{\underline{8}} \left\langle\lambda_{23} 
L_\mu L^\mu \right\rangle\;+\; {\rm h.c.},\nonumber \\
{\cal L}^{(1/2)}_{\underline{27}} &=& 
-\frac{G_F}{\sqrt{2}} V_{ud} V^*_{us}f^4_{\pi}
g_{\underline{27}}\left(\left\langle \lambda_{12} L_\mu\right\rangle
\left\langle \lambda_{31} L^\mu \right\rangle
\right. \nonumber \\ && \hspace{1.5cm} \left.  -
\left\langle \lambda_{32} L_\mu\right\rangle
\left\langle \lambda_{11} L^\mu \right\rangle 
- 5 \left\langle \lambda_{32} L_\mu\right\rangle
\left\langle \lambda_{33} L^\mu \right\rangle\right) +\; {\rm h.c.}, 
\nonumber \\
{\cal L}^{(3/2)}_{\underline{27}} &=& 
-\frac{G_F}{\sqrt{2}} V_{ud} V^*_{us}f^4_{\pi}
g_{\underline{27}}\left(\left\langle \lambda_{12} L_\mu\right\rangle
\left\langle \lambda_{31} L^\mu \right\rangle 
\right. \nonumber \\ && \hspace{1.5cm} \left.  +2 
\left\langle \lambda_{32} L_\mu\right\rangle
\left\langle \lambda_{11} L^\mu \right\rangle 
+ \left\langle \lambda_{32} L_\mu\right\rangle
\left\langle \lambda_{33} L^\mu \right\rangle\right) +\; {\rm h.c.}
\label{Eq:L8L27}
\end{eqnarray}
The coupling constants $g_{\underline{8}}$ and $g_{\underline{27}}$ 
can be extracted from the $K\rightarrow \pi\pi$ decay rate 
and the $\Delta T = 1/2$ enhancement is reflected in these constants.

Now, we are in a position to evaluate the constants
 $g_{\underline{8}}$ and $g_{\underline{27}}$
 from the results of $\langle {\cal Q}_i \rangle$.
Comparison of  Eq.(\ref{Eq:part1}) and Eqs.(\ref{Eq:qa}-\ref{Eq:qz})
with Eqs.(\ref{Eq:Lcpt2},\ref{Eq:L8L27}) yields the following results:
\begin{eqnarray}
g^{(1/2)}_{\underline{8}} &=& -\frac25 c_1 + \frac35 c_2 + c_4 
+ \left(\frac{\langle \bar{q} q \rangle}{Mf^2_{\pi}}
-\frac{\langle \bar{q} q \rangle N_c M}{8\pi^2 f^4_{\pi}}\right)c_6 
-\frac35 c_9 + \frac25 c_{10}, \nonumber \\ 
g^{(1/2)}_{\underline{27}} &=& 
\frac{1}{15} c_1 + \frac{1}{15} c_2 + \frac{1}{10} c_9 
+ \frac{1}{10} c_{10},\nonumber \\
g^{(3/2)}_{\underline{27}} &=& 5\  g^{(1/2)}_{\underline{27}}.
\label{Eq:ratiol}
\end{eqnarray}
We employed the Wilson coefficients $c_i$ obtained by Buchalla {\em et al.} 
~\cite{Burasetal} 
as shown in Table I.  There are three different
renormalization schemes.  The LO denotes the summation of the leading
logarithmic terms $\sim \alpha_{\rm s} \ln (M_W/\mu))^n$, which were
mainly done by Vainshtein {\em et al.}~\cite{VZS,SVZ}, Gilman and
Wise~\cite{GW} and Guberina and Peccei~\cite{GP}.  The NDR and HV 
represent respectively ``Naive dimensional regularization'' and
't Hooft-Veltman scheme~\cite{HV,BM} (see Ref.~\cite{Burasetal}
for details).  In the leading order contribution in the large $N_c$ 
expansion three parameters are involved: the pion decay constant, 
the quark condensate, and the constituent quark mass.  The values of 
the quark condensate and constituent quark mass 
are the parameters we can play with.  However, these two parameters are 
to some extent theoretically restricted.  The value of the quark condensate
lies between 
$-(300 \ {\rm MeV})^3 \leq \langle \bar{q} q\rangle/2
\leq -(200 \ {\rm MeV})^3$.
  Larger values give slightly better ratio of
the constants $g_{\underline{8}}$ and $g_{\underline{27}}$.  The 
constituent quark mass is in fact the free parameter of the $\chi$QM.  It 
is known that the value $M\simeq 400$ MeV describes consistently very well
the static properties of the baryon~\cite{review}.  However, 
in the mesonic sector lower values are often voted
\cite{BPKG}.  We also find that lower values of the constituent
quark mass provide better ratios of the constants.  Figure 1 shows the 
dependence of the $g_{\underline{8}}/g_{\underline{27}}$ on the $M$.  In 
the NDR scheme $M$-dependence is stronger than in the other two 
schemes.  One can easily understand this dependence.  The parameter 
$M$ appears in front of the coefficient $c_6$ in 
Eq.(\ref{Eq:ratiol}).  From Table I we find that the Wilson 
coefficient $c_6$ based on the NDR scheme ($-0.0022$)
is larger than in the other two schemes ($-0.009$) which causes the 
strong dependence of the ratio $g_{\underline{8}}/g_{\underline{27}}$ 
on the $M$ in the case of the NDR scheme.  Because of the same reason,
its dependence on the quark condensate looks very similar, 
see Fig. 2.  In Table II we find that the ratio 
$g_{\underline{8}}/g_{\underline{27}}$ is almost seven times underestimated,
compared to the empirical data ($g_{\underline{8}}/g_{\underline{27}}
\simeq 22$ with the counterterms).    

From the calculation of the $\langle {\cal Q}_{\Delta S = 2}\rangle$
in Eq.(\ref{Eq:qz}), we easily write the effective $\Delta S = 2$ weak 
chiral Lagrangian to order ${\cal O}(p^2)$:
\begin{equation}
{\cal L}^{\Delta S=2,{\cal O}(p^4)}_{\rm eff} 
\;=\; -\frac{G^2_{F} M^2_{W}}{4\pi^2}
{\cal F} \left(\lambda_c, \lambda_t,m^2_{c}, m^2_{t}, M^2_{W}
\right)b(\mu) f^4_{\pi} \langle 
\lambda_6 L_\mu\rangle\langle \lambda_6 L^\mu\rangle.
\end{equation}

\subsection{${\cal O} (N_c)$ and axial-vector coupling corrections}
So far we concentrate on the leading order in the large $N_c$ expansion.
We now want to introduce the next-to-leading order corrections in 
the large $N_c$ expansion.  The ${\cal Q}_{3}$ and ${\cal Q}_5$ survive 
and the additional terms come into existence in the other 
quark operators.  The VEV of the quark operators are obtained in the 
${\cal O}(N_c)$ order as follows:
\begin{eqnarray}
{\langle Q_1+{\cal Q}^{\dagger}_1 \rangle}^{{\cal O}
\left( N_c \right)}_{{\cal O} (p^2)} &=&
{ f_\pi^4 \over N_c } 
\left( {3 \over 5} \langle \lambda_6 L_\mu L_\mu  \rangle
+{1 \over 3}  t_{ij;kl}
\langle \lambda_{ij} L_\mu  \rangle \langle \lambda_{kl} L_\mu 
\rangle \right) \\
{\langle Q_2 +{\cal Q}^{\dagger}_2\rangle}^{{\cal O}
\left( N_c \right)}_{{\cal O} (p^2)} &=&
{ f_\pi^4 \over N_c } 
\left( -{2 \over 5} \langle \lambda_6 L_\mu L_\mu  \rangle
+ {1 \over 3} t_{ij;kl}
\langle \lambda_{ij} L_\mu  \rangle \langle \lambda_{kl} L_\mu 
\rangle \right) \\
{\langle Q_3 +{\cal Q}^{\dagger}_3\rangle}^{{\cal O}
\left( N_c \right)}_{{\cal O} (p^2)} &=&
{f_\pi^4 \over N_c } 
\langle \lambda_6 L_\mu L_\mu  \rangle \\
{\langle Q_4 +{\cal Q}^{\dagger}_4\rangle}^{{\cal O}
\left( N_c \right)}_{{\cal O} (p^2)} &=&
 0 \\
{\langle Q_5 +{\cal Q}^{\dagger}_5\rangle}^{{\cal O}
\left( N_c \right)}_{{\cal O} (p^2)} &=&
\left(
{ f_\pi^2 \langle \bar{q} q \rangle \over N_c M} - 
{\langle \bar{q} q \rangle M \over 8 \pi^2} \right) 
\langle \lambda_6 L_\mu L_\mu  \rangle \\
{\langle Q_6 +{\cal Q}^{\dagger}_6\rangle}^{{\cal O}
\left( N_c \right)}_{{\cal O} (p^2)} &=&
 0 \\
\langle {\cal Q}_7+{\cal Q}^{\dagger}_7\rangle_{{\cal O}(p^2)}^{{\cal O}(N_c)}
&=& \left( \frac{3 f^2_\pi \langle \bar{q} q \rangle}{4 N_c M}
- \frac{3 \langle \bar{q} q \rangle M}{32 \pi^2} \right)
\left( \langle \lambda_6 \partial_\mu U^{\dagger} \partial_\mu U U^{\dagger}
\hat{Q} U \rangle
+ \langle \lambda_6 U^{\dagger} \hat{Q} U \partial_\mu U^{\dagger} \partial_\mu
U \rangle \right) \\
\langle {\cal Q}_8+{\cal Q}^{\dagger}_8\rangle_{{\cal O}(p^2)}^{{\cal O}(N_c)}
&=& \frac{3 f^4_\pi}{2 N_c} \, \langle \lambda_6 U^{\dagger} \partial_\mu U
\rangle \langle \hat{Q} \partial_\mu U U^{\dagger} \rangle \\
\nonumber \\
{\langle Q_9 +{\cal Q}^{\dagger}_9\rangle}^{{\cal O}
\left( N_c \right)}_{{\cal O} (p^2)} &=&
{ f_\pi^4 \over N_c} 
\left( {2 \over 5} \langle \lambda_6 L_\mu L_\mu \rangle
+{1 \over 2} t_{ij;kl}
\langle \lambda_{ij} L_\mu  \rangle \langle \lambda_{kl} L_\mu 
\rangle \right) \\
{\langle Q_{10} +{\cal Q}^{\dagger}_{10}
\rangle}^{{\cal O}\left( N_c \right)}_{{\cal O} (p^2)} &=&
{ f_\pi^4 \over N_c } 
\left( -{3 \over 5} \langle \lambda_6 L_\mu L_\mu  \rangle
+{1 \over 2} t_{ij;kl}
\langle \lambda_{ij} L_\mu  \rangle \langle \lambda_{kl} L_\mu 
\rangle \right) .
\end{eqnarray}
Taking into account the ${\cal O} (N_c)$ corrections given above, 
we get the $g_{\underline{8}}$ and $g_{\underline{27}}$:
\begin{eqnarray}
g_{\underline{8}}^{\left( {\cal O}\left( N_c^2 \right) 
+ {\cal O}\left( N_c \right) \right)} &=& 
 \left( -\frac{2}{5} + {1 \over N_c} \frac{3}{5} \right) c_1 
+ \left(  \frac{3}{5} - {1 \over N_c} \frac{2}{5} \right) c_2 
+ {1 \over N_c} c_3 + c_4
\nonumber \\
&& + \left( 
 {\langle \bar{q} q\rangle \over N_c f_\pi^2 M} 
- {\langle \bar{q} q\rangle M \over 8 f_\pi^4 \pi^2} 
\right) c_5
+ \left( 
{\langle \bar{q} q\rangle \over f_\pi^2 M} 
- {N_c \langle \bar{q} q\rangle M \over 8 f_\pi^4 \pi^2} 
 \right) c_6 \nonumber \\
&& + \left( - \frac{3}{5} + {1 \over N_c} \frac{2}{5} \right) c_9 
+ \left(  \frac{2}{5} - {1 \over N_c} \frac{3}{5} \right) c_{10} 
\label{Eq:g8nlo} \\ 
g_{\underline{27}}^{\left( {\cal O}\left( N_c^2 \right)
+ {\cal O}\left( N_c \right) \right)} &=& 
\left(1 + {1 \over N_c} \right) \left( \frac{3}{5} c_1 
+ \frac{3}{5} c_2 + \frac{9}{10} c_9 + \frac{9}{10} c_{10}\right).
\label{Eq:g27nlo}
\end{eqnarray}
Because of the sign in the $1/N_c$ corrections in 
Eqs.(\ref{Eq:g8nlo},\ref{Eq:g27nlo}), we can easily see that
the ${\cal O}(N_c)$ correction suppresses the octet coupling 
$g_{\underline{8}}$ while enhancing the eikosiheptaplet coupling 
$g_{\underline{27}}$.  It indicates that the ${\cal O}(N_c)$
corrections make the ratio of these two couplings even worse
than that with only the leading contribution.  As shown in Table III
the ratio $g_{\underline{8}}/g_{\underline{27}}$ is completely 
underestimated.  

It is also interesting to consider the effect of the quark axial-vector 
coupling constant.  To be more consistent in the large $N_c$ expansion,
we can take into account subleading order couplings in the large $N_c$
in addition to the leading order coupling given by 
$\bar{\psi} U^{\gamma_5} \psi$.  The simplest way of generalizing the
$\chi$QM is to introduce the quark axial-vector coupling $g_A$ 
different from unity~\cite{JJS,Prasz}. In such a case the $g_A$ is known 
to be
smaller than 1. The $g_A$ enters in the effective action 
given in Eq.(\ref{Eq:echa}):
\begin{equation}
S_{\rm eff} [\pi] \;=\; -N_c {\rm Tr} \ln \left(i\rlap{/}{\partial} 
+ iM U^{\gamma_5} + i \epsilon_A U^{\gamma_5}
\rlap{/}{\partial} U^{\gamma_5} \right),
\label{Eq:actionga}
\end{equation}  
where $\epsilon_A = (1-g_A)/2$.  The understanding of this coupling
depends on the specific dynamical assumptions.  There are two different
arguments about the large $N_c$ behavior of the $1-g_A^{2}$.  For 
example, Weinberg argued that $1-g_A^{2}$ is of order ${\cal O} (1/N_c)$
using the Adler-Weisberger sum rule~\cite{Weinberg}. Also  Dicus {\em et 
al.}~\cite{Dicusetal} considered it as $1/N_c$ 
corrections.  On the other hand, Broniowski {\em et al.}~\cite{BSL}
demonstrated that from the Adler-Weisberger sum 
rule with the reggeized $\rho$ meson exchange $1-g_A^{2}$ is of 
order ${\cal O} (N^{0}_c)$.  The new term $i \epsilon_A U^{\gamma_5}
\rlap{/}{\partial} U^{\gamma_5}$ being considerd, the operators
${\left( A_{1,2} \right)}_\mu$ can be rewritten as
\begin{eqnarray}
{\left( A_{1,2} \right)}_\mu &=& \int \frac{d^4 k}{{\left(2 \pi \right)}^4}\,
 {1 \over k^2 + M^2} \sum_{n=0}^{\infty}\left[ \left(
\partial^2 - 2 i k \partial + M \left(
\rlap{/}{\partial} U^{\gamma_5}\right) 
\; + \; \epsilon_A  \left( \left(\rlap{/}{\partial} 
U^{\gamma_5}\right)\left(\rlap{/}{\partial} 
U^{\gamma_5}\right)\right.\right.\right.\nonumber \\
&&+ U^{- \gamma_5} \left(\partial^2 U^{\gamma_5} \right)
+ 2 U^{- \gamma_5} \left(\partial_\nu  U^{\gamma_5} \right) \partial_\nu
+ 2 \left(\rlap{/}{\partial} U^{\gamma_5}\right) 
U^{- \gamma_5} \rlap{/}{\partial} \nonumber \\ && 
- 2  i  U^{- \gamma_5} \left(\partial_\nu  U^{\gamma_5} \right) k_\nu
- 2  i \left(\rlap{/}{\partial} U^{\gamma_5}) U^{- \gamma_5} \rlap{/}{k}
\right) \nonumber \\ &&\left.\left.
+ \epsilon_A^2 \left(\rlap{/}{\partial} U^{\gamma_5}\right) 
\left(\rlap{/}{\partial} U^{\gamma_5}\right) \right) 
{1 \over k^2 + M^2} \right]^n \nonumber \\
&& \times\left(  i\rlap{/}{\partial}
  -  i\, M U^{-\gamma_5} +  i \,
\epsilon_A \left(\rlap{/}{\partial} U^{\gamma_5}\right)
U^{- \gamma_5} + \rlap{/}{k} \right)
 \gamma_{\mu} P_{R,L} \Lambda_{1,2} \; .
\end{eqnarray}   
Since the parameter $\epsilon_{\rm A}$ is tiny, we can safely neglect the
$\epsilon_{\rm A}^2$ terms.  Thus, we obtain the following results:
\begin{eqnarray}
{\langle {\cal Q}_1 +{\cal Q}^{\dagger}_1 
\rangle}^{{\cal O}(\epsilon_{\rm A})}_{{\cal O}(p^2)} 
&=& -f_\pi^4\epsilon_A \left( { \langle \bar{q} q \rangle 
\over 2 M f_\pi^2} + 3 \right) 
\nonumber \\ &\times&
\left(-\frac25 \langle \lambda_6 L_\mu L_\mu 
\rangle + \frac13 t_{ij;kl} \langle \lambda_{ij} L_\mu \rangle\langle 
\lambda_{kl} L_\mu \rangle\right), \\
{\langle {\cal Q}_2 +{\cal Q}^{\dagger}_2 
\rangle}^{{\cal O}(\epsilon_{\rm A})}_{{\cal O}(p^2)} 
&=& -f_\pi^4\epsilon_A \left( {\langle \bar{q} q \rangle 
\over 2 M f_\pi^2} + 3 \right) 
\nonumber \\ &\times&
\left(\frac35 \langle \lambda_6 L_\mu L_\mu 
\rangle + \frac13 t_{ij;kl} \langle \lambda_{ij} L_\mu \rangle\langle 
\lambda_{kl} L_\mu \rangle\right),\\
{\langle {\cal Q}_3 + {\cal Q}^{\dagger}_3 
\rangle}^{{\cal O}(\epsilon_{\rm A})}_{{\cal O}(p^2)} &=& 0 \\
{\langle {\cal Q}_4 + {\cal Q}^{\dagger}_4 
\rangle}^{{\cal O} (\epsilon_{\rm A})}_{{\cal O}(p^2)} &=& 
-f_\pi^4\epsilon_A \left( {\langle \bar{q} q \rangle 
\over 2 M f_\pi^2} + 3 \right) 
\langle \lambda_6  L_\mu  L_\mu \rangle,\\
{\langle {\cal Q}_5 + {\cal Q}^{\dagger}_5
\rangle}^{{\cal O}(\epsilon_{\rm A})}_{{\cal O}(p^2)} &=& 0 \\
{\langle {\cal Q}_6 +{\cal Q}^{\dagger}_6 
\rangle}^{{\cal O} (\epsilon_{\rm A})}_{{\cal O}(p^2)} &=&
-f_\pi^4 \epsilon_A \left({ 4 \langle \bar{q} q \rangle \over M f_\pi^2 } 
- {3 N_c \langle \bar{q} q \rangle M \over 4 \pi^2 f_\pi^2 }\right) 
\langle L_\mu L_\mu \lambda_6 \rangle, \\
\langle {\cal Q}_7+{\cal Q}^{\dagger}_7\rangle_{{\cal O}(p^2)}^{{\cal O}
(\epsilon_A)}
&=& - \epsilon_A \left( \frac{3 f^2_\pi \langle \bar{q} q \rangle}{4 M} 
+ \frac{9}{2} f^4_\pi \right) 
\langle \lambda_6 U^{\dagger} \partial_\mu U
\rangle \langle \hat{Q} \partial_\mu U U^{\dagger} \rangle \\
\langle {\cal Q}_8+{\cal Q}^{\dagger}_8\rangle_{{\cal O}(p^2)}^{{\cal O}
(\epsilon_A)}
&=& - \epsilon_A \left( \frac{3 f^2_\pi \langle \bar{q} q \rangle}{M}
- \frac{9 N_c \langle \bar{q} q \rangle M}{16 \pi^2} \right), \\
&& \times 
\left( \langle \lambda_6 \partial_\mu U^{\dagger} \partial_\mu U U^{\dagger}
\hat{Q} U \rangle
+ \langle \lambda_6 U^{\dagger} \hat{Q} U \partial_\mu U^{\dagger} 
\partial_\mu U \rangle \right),\\
{\langle {\cal Q}_9 +{\cal Q}^{\dagger}_9
\rangle}^{{\cal O}(\epsilon_{\rm A})}_{{\cal O}(p^2)} &=& 
- f_\pi^4\epsilon_A \left( {\langle \bar{q} q \rangle \over 
2 M f_\pi^2} + 3 \right) 
\nonumber \\ &\times&
\left(-\frac35 \langle \lambda_6 L_\mu L_\mu 
+ \frac12 t_{ij;kl} \langle \lambda_{ij} L_\mu \rangle\langle 
 \lambda_{kl} L_\mu \rangle\right), \\
{\langle {\cal Q}_{10} +{\cal Q}^{\dagger}_{10}
\rangle}^{{\cal O}(\epsilon_{\rm A})}_{{\cal O}(p^2)} 
&=& - f_\pi^4 \epsilon_A 
\left( {\langle \bar{q} q \rangle \over 2 M f_\pi^2} + 3 \right) 
\nonumber \\ &\times&
\left(\frac25 \langle \lambda_6 L_\mu L_\mu 
+ \frac12 t_{ij;kl} \langle \lambda_{ij} L_\mu \rangle\langle 
 \lambda_{kl} L_\mu \rangle\right).
\end{eqnarray}
The LECs can be then obtained as follows:
\begin{eqnarray}
g_{\underline{8}}^{g_A} &=&
\left( 1 - \epsilon_A \left( { \langle \bar{q} q \rangle 
\over 2 M f_\pi^2} + 3 \right) \right)
\left( - \frac{2}{5} c_1 + \frac{3}{5} c_2 + c_4
- \frac{3}{5} c_9 + \frac{2}{5} c_{10} \right)
\nonumber \\ &&
+ \left( {\langle \bar{q} q \rangle \over f_\pi^2 M} - 
{N_c \langle \bar{q} q \rangle M \over 8 f_\pi^4 \pi^2}
- \epsilon_A \left(
{ 4 \langle \bar{q} q \rangle \over M f_\pi^2 } 
- {3 N_c \langle \bar{q} q \rangle M \over 4 \pi^2 f_\pi^2 }
\right) \right) c_6
\nonumber \\ 
g_{\underline{27}}^{g_A} &=&
\left( 1 - \epsilon_A \left( { \langle \bar{q} q \rangle 
\over 2 M f_\pi^2} + 3 \right) \right)
\left(
\frac{3}{5} c_1 + \frac{3}{5} c_2 + \frac{9}{10} c_9 + \frac{9}{10} c_{10}
\right).
\label{Eq:ratioa}
\end{eqnarray}
Figure 3 shows the dependence of the 
$g_{\underline{8}}/g_{\underline{27}}$ on the
quark axial-vector constant $g_{\rm A}$ ranging from $0.75$ to $1.25$.  The 
dependence on the $g_{\rm A}$ is stronger again in the case of the 
NDR scheme.  To get a reasonable value for the ratio one should choose a
 large value of $g_{\rm A}$ 
which, however, deviates from the physical value $g_{\rm A}\simeq 
0.75$, 
as easily found from Eq.(\ref{Eq:ratioa}).  Thus, corrections from 
the ${\cal O}(N_c)$ and axial-vector coupling constants turn out to be quite
useless if one wants to reproduce the empirical data.    

The effective $\Delta S = 2$ weak chiral Lagrangian with the $1/N_c$ 
and $g_{\rm A}$ corrections is given as follows:
\begin{eqnarray}
{\cal L}^{\Delta S=2,{\cal O}(p^2)}_{\rm eff} 
&=& -\frac{G^2_{F} M^2_{W}}{4\pi^2}
{\cal F} \left(\lambda_c, \lambda_t,m^2_{c},
m^2_{t}, M^2_{W}\right)b(\mu) \nonumber \\
&&\hspace{0.8cm} \left[f^4_{\pi} + \frac{f^4_\pi}{N_c} 
- \epsilon_A \left( \frac{f^2_\pi \langle \bar{q} q \rangle}{2 M}
+ 3 f^4_\pi \right)\right]
\langle \lambda_6 L_\mu\rangle\langle \lambda_6 L^\mu\rangle.
\end{eqnarray}
\section{Next-to-leading order ${\cal O} (p^4)$}
Although the derivative expansion to order ${\cal O} (p^4)$ 
is straightforward, reducing the number of terms is quite involved.  
We first can reduce the terms containing higher-order derivatives
by using the following identities:
\begin{eqnarray}
U^+ \left( \partial_\mu \partial_\nu U \right) &=&
- {1 \over 2} \left\{ L_\mu,L_\nu \right\} - {1 \over 4}  i\, W_{\mu \nu}\, , \\
\left( \partial_\mu \partial_\nu U^+ \right) U &=&
- {1 \over 2} \left\{ L_\mu,L_\nu \right\} + {1 \over 4}
 i\, W_{\mu \nu} \, ,
\end{eqnarray}
where
\begin{equation}
W_{\mu \nu} = 2 \left( \partial_\mu L_\nu + \partial_\nu  L_\mu \right) \, .
\end{equation}
We can then compare the reduced set of terms in the ${\cal O} (p^4)$
order Lagrangian with that in Ref.~\cite{Kamboretal}.  To this end the 
number of terms can be reduced further by employing the 
equation of motion for the meson fields in the chiral limit
$\partial_\mu L_\mu=0$ and the identities 
\begin{eqnarray}
\label{Eq:id1}
{ 1 \over 8 } \langle W_{\mu \nu}^2 \Lambda \rangle &=&
\langle L_\mu L_\nu L_\nu  L_\mu \Lambda \rangle
- \langle L_\mu L_\nu L_\mu L_\nu \Lambda \rangle, \\
{1 \over 4}  i\, \langle L_\mu \Lambda \rangle \langle
\left[ W_{\mu \nu} , L_\nu \right]
\Lambda \rangle &=&
\langle L_\mu L_\nu \Lambda \rangle \langle L_\nu L_\mu \Lambda \rangle
- \langle L_\mu L_\nu \Lambda \rangle \langle L_\mu L_\nu \Lambda \rangle
\nonumber \\
&& + \langle L_\mu \Lambda \rangle \langle L_\nu L_\mu L_\nu \Lambda \rangle
- {1 \over 2} \langle L_\mu \Lambda \rangle \langle
\left\{ L_\mu , L_\nu  L_\nu \right\}
\Lambda \rangle, \\                                 
{1 \over 2}  i\, \langle L_\mu \Lambda \rangle
\langle W_{\mu \nu} L_\nu  \rangle &=&
\langle L_\mu L_\nu \Lambda \rangle \langle L_\nu L_\mu \Lambda \rangle
- \langle L_\mu L_\nu \Lambda \rangle \langle L_\mu L_\nu \Lambda \rangle
\nonumber \\
&& + \langle L_\mu \Lambda \rangle \langle L_\nu L_\mu L_\nu \Lambda \rangle
- \langle L_\mu \Lambda \rangle \langle L_\nu L_\nu L_\mu \Lambda \rangle\, ,
\label{Eq:id2}
\end{eqnarray}
where $\Lambda$ denote arbitrary flavor matrices.
The identities (\ref{Eq:id1})-(\ref{Eq:id2}) can be easily obtained
by integration by parts and some trace identities from
the Cayley-Hamilton theorem.  Decomposing the octet and eikosiheptaplet 
contributions, we end up with the following results for the vacuum  expectation 
values at ${\cal O}(p^4)$ order and leading order in the large $N_c$ expansion:
\begin{eqnarray} 
\label{Eq:nlo1}
{\langle {\cal Q}_1 + {\cal Q}^\dagger_1 \rangle}_{{\cal O}(p^4)}
 &=&  {N_c f_\pi^2 \over 24 \pi^2}
 \Big[
{4 \over 5} \langle L_\mu L_\nu L_\nu L_\mu \lambda_6 \rangle
- {3 \over 5} \langle L_\mu \lambda_6 \rangle
\langle L_\mu L_\nu L_\nu \rangle
\nonumber \\ && \hspace{15mm}
- {3 \over 5} \varepsilon_{\alpha \beta \gamma \delta}
\langle L_\alpha \lambda_6 \rangle \langle L_\beta L_\gamma L_\delta \rangle
\nonumber \\
&& \hspace{15mm} - {2 \over 3} t_{ijkl} \Big(
\langle L_\mu L_\nu \lambda_{ij} \rangle
\langle L_\nu L_\mu \lambda_{kl} \rangle
- \langle L_\mu L_\nu \lambda_{ij} \rangle
\langle L_\mu L_\nu \lambda_{kl} \rangle
\nonumber \\
&& \hspace{25mm}
+ \langle L_\mu \lambda_{ij} \rangle
\langle L_\nu L_\mu L_\nu \lambda_{kl} \rangle
+ \varepsilon_{\alpha \beta \gamma \delta}
\langle L_\alpha \lambda_{ij} \rangle \langle L_\beta L_\gamma L_\delta
\lambda_{kl} \rangle \Big) \Big] \\
{\langle {\cal Q}_2 + {\cal Q}^\dagger_2 \rangle}_{{\cal O}(p^4)} 
 &=&  {N_c f_\pi^2 \over 24 \pi^2}
\Big[
- {6 \over 5} \langle L_\mu L_\nu L_\nu L_\mu \lambda_6 \rangle
+ {2 \over 5} \langle L_\mu \lambda_6 \rangle
\langle L_\mu L_\nu L_\nu \rangle
\nonumber \\ && \hspace{15mm}
+ {2 \over 5} \varepsilon_{\alpha \beta \gamma \delta}
\langle L_\alpha \lambda_6 \rangle \langle L_\beta L_\gamma L_\delta \rangle
\nonumber \\
&& \hspace{15mm} - {2 \over 3} t_{ijkl} \Big(
\langle L_\mu L_\nu \lambda_{ij} \rangle
\langle L_\nu L_\mu \lambda_{kl} \rangle
- \langle L_\mu L_\nu \lambda_{ij} \rangle
\langle L_\mu L_\nu \lambda_{kl} \rangle
\nonumber \\
&& \hspace{25mm}
+ \langle L_\mu \lambda_{ij} \rangle
\langle L_\nu L_\mu L_\nu \lambda_{kl} \rangle
+ \varepsilon_{\alpha \beta \gamma \delta}
\langle L_\alpha \lambda_{ij} \rangle \langle L_\beta L_\gamma L_\delta
\lambda_{kl} \rangle \Big) \Big] \\                                  
{\langle {\cal Q}_3 + {\cal Q}^\dagger_3 \rangle}_{{\cal O}(p^4)} 
 &=& - {N_c f_\pi^2 \over 24 \pi^2}
\Big[
\langle L_\mu \lambda_6 \rangle
\langle L_\mu L_\nu L_\nu \rangle
+ \varepsilon_{\alpha \beta \gamma \delta}
\langle L_\alpha \lambda_6 \rangle \langle L_\beta L_\gamma L_\delta \rangle
\Big] \\
{\langle {\cal Q}_4 + {\cal Q}^\dagger_4  \rangle}_{{\cal O}(p^4)}  
&=& - {N_c f_\pi^2 \over 12 \pi^2}
\langle L_\mu L_\nu L_\nu L_\mu \lambda_6 \rangle \\
{\langle {\cal Q}_5 + {\cal Q}^\dagger_5 \rangle}_{{\cal O}(p^4)} 
 &=& {N_c f_\pi^2 \over 24 \pi^2}
\Big[ 
\langle L_\mu \lambda_6 \rangle
\langle L_\mu L_\nu L_\nu \rangle
- \varepsilon_{\alpha \beta \gamma \delta}
\langle L_\alpha \lambda_6 \rangle \langle L_\beta L_\gamma L_\delta \rangle
\Big] \\                 
{\langle {\cal Q}_6 + {\cal Q}^\dagger_6 \rangle}_{{\cal O}(p^4)} 
 &=&
\left( {N_c^2 M^2 \over 128 \pi^4} - {N_c f_\pi^2 \over 8 \pi^2}
+ {f_\pi^4 \over 2 M^2} \right)
\langle L_\mu L_\mu L_\nu L_\nu \lambda_6 \rangle \\
\langle {\cal Q}_7+{\cal Q}^{\dagger}_7\rangle_{{\cal O}(p^4)}
&=& - \frac{3 N_c f^2_\pi}{48 \pi^2} \, \Big[
2 \langle \lambda_6 \partial_\mu U^{\dagger} \partial_\nu U \rangle
\langle \hat{Q} \partial_\nu U \partial_\mu U^{\dagger} \rangle
-2 \langle \lambda_6 \partial_\mu U^{\dagger} \partial_\nu U \rangle
\langle \hat{Q} \partial_\mu U \partial_\nu U^{\dagger} \rangle
\nonumber \\
&& \hspace{-20mm}
+ \langle \lambda_6 U^{\dagger} \partial_\mu U \rangle
\langle \hat{Q} \partial_\nu U \partial_\mu U^{\dagger} \partial_\nu U
U^{\dagger} \rangle
+ \langle \lambda_6 U^{\dagger} \partial_\nu U \partial_\mu U^{\dagger}
\partial_\nu U \rangle \langle \hat{Q} \partial_\mu U U^{\dagger} \rangle
\nonumber \\
&& \hspace{-20mm}
- \varepsilon_{\alpha \beta \gamma \delta}
\left(
\langle \lambda_6 U^{\dagger} \partial_\alpha U \rangle
\langle \hat{Q} \partial_\beta U \partial_\gamma U^{\dagger} \partial_\delta U
U^{\dagger} \rangle
+ \langle \lambda_6 U^{\dagger} \partial_\alpha U \partial_\beta U^{\dagger}
\partial_\gamma U \rangle \langle \hat{Q} \partial_\delta U U^{\dagger} \rangle
\right) \Big]
\\   
\langle {\cal Q}_8+{\cal Q}^{\dagger}_8\rangle_{{\cal O}(p^4)}^{{\cal O}(N_c^2)}
&=& \left(\frac{3 N_c^2 M^2}{256 \pi^4} - \frac{3 N_c f^2_\pi}{16 \pi^2}
+ \frac{3 f^4_\pi}{4 M^2} \right) \,
\langle \lambda_6 \partial_\mu U^{\dagger} \partial_\mu U U^{\dagger}
\hat{Q} U \partial_\nu U^{\dagger} \partial_\nu U \rangle
\nonumber \\
&& \hspace{-20mm}
+ \frac{3 N_c \langle \bar{q} q \rangle}{192 M \pi^2} \Big[
\langle \lambda_6 \partial_\mu U^{\dagger} \partial_\nu U
\partial_\mu U^{\dagger} \partial_\nu U U^{\dagger} \hat{Q} U \rangle
+ \langle \lambda_6 U^{\dagger} \hat{Q} U \partial_\mu U^{\dagger}
\partial_\nu U \partial_\mu U^{\dagger} \partial_\nu  U \rangle
\nonumber \\
&& \hspace{-20mm}
- \langle \lambda_6 \partial_\mu U^{\dagger} \partial_\nu U
\partial_\nu U^{\dagger} \partial_\mu U U^{\dagger} \hat{Q} U \rangle
- \langle \lambda_6 U^{\dagger} \hat{Q} U \partial_\mu U^{\dagger}
\partial_\nu U \partial_\nu U^{\dagger} \partial_\mu  U \rangle
\nonumber \\
&& \hspace{-20mm}
+ \frac{1}{2} i \langle \lambda_6 \left\{ \partial_\mu U^{\dagger}
\partial_\nu  U , W_{\mu \nu} \right\} U^{\dagger} \hat{Q} U \rangle
- \frac{1}{2} i \langle \lambda_6 U^{\dagger} \hat{Q}
U \left\{ \partial_\mu U^{\dagger}
\partial_\nu  U , W_{\mu \nu} \right\} \rangle
\nonumber \\
&& \hspace{-20mm}
+ \frac{1}{8} \langle \lambda_6 W_{\mu \nu}^2 U^{\dagger} \hat{Q} U \rangle
+ \frac{1}{8} \langle \lambda_6 U^{\dagger}  \hat{Q} U W_{\mu \nu}^2 \rangle
\nonumber \\
&& \hspace{-20mm}
- 2 \varepsilon_{\alpha \beta \gamma \delta} \left(
\langle \lambda_6 \partial_\alpha U^{\dagger} \partial_\beta U
\partial_\gamma  U^{\dagger} \partial_\delta U  U^{\dagger} \hat{Q} U \rangle
- \langle \lambda_6 U^{\dagger}
\hat{Q} U \partial_\alpha U^{\dagger} \partial_\beta U
\partial_\gamma  U^{\dagger} \partial_\delta U \rangle \right)
\Big]                                    
\\        
{\langle {\cal Q}_9 + {\cal Q}^\dagger_9 \rangle}_{{\cal O}(p^4)} 
 &=&  {N_c f_\pi^2 \over 24 \pi^2}
\Big[
 {6 \over 5} \langle L_\mu L_\nu L_\nu L_\mu \lambda_6 \rangle
- {2 \over 5} \langle L_\mu \lambda_6 \rangle
\langle L_\mu L_\nu L_\nu \rangle
\nonumber \\ && \hspace{15mm}
- {2 \over 5} \varepsilon_{\alpha \beta \gamma \delta}
\langle L_\alpha \lambda_6 \rangle \langle L_\beta L_\gamma L_\delta \rangle
\nonumber \\
&& \hspace{15mm} -  t_{ijkl} \Big(
\langle L_\mu L_\nu \lambda_{ij} \rangle
\langle L_\nu L_\mu \lambda_{kl} \rangle
- \langle L_\mu L_\nu \lambda_{ij} \rangle
\langle L_\mu L_\nu \lambda_{kl} \rangle
\nonumber \\
&& \hspace{25mm}
+ \langle L_\mu \lambda_{ij} \rangle
\langle L_\nu L_\mu L_\nu \lambda_{kl} \rangle
+ \varepsilon_{\alpha \beta \gamma \delta}
\langle L_\alpha \lambda_{ij} \rangle \langle L_\beta L_\gamma L_\delta
\lambda_{kl} \rangle \Big) \Big] \\
{\langle {\cal Q}_{10} + {\cal Q}^\dagger_{10} \rangle}_{{\cal O}(p^4)}  &=& 
 {N_c f_\pi^2 \over 24 \pi^2}
\Big[
-{4 \over 5} \langle L_\mu L_\nu L_\nu L_\mu \lambda_6 \rangle
+ {3 \over 5} \langle L_\mu \lambda_6 \rangle
\langle L_\mu L_\nu L_\nu \rangle
\nonumber \\ && \hspace{15mm}
+ {3 \over 5} \varepsilon_{\alpha \beta \gamma \delta}
\langle L_\alpha \lambda_6 \rangle \langle L_\beta L_\gamma L_\delta \rangle
\nonumber \\
&& \hspace{15mm} - t_{ijkl} \Big(
\langle L_\mu L_\nu \lambda_{ij} \rangle
\langle L_\nu L_\mu \lambda_{kl} \rangle
- \langle L_\mu L_\nu \lambda_{ij} \rangle
\langle L_\mu L_\nu \lambda_{kl} \rangle
\nonumber \\
&& \hspace{25mm}
+ \langle L_\mu \lambda_{ij} \rangle
\langle L_\nu L_\mu L_\nu \lambda_{kl} \rangle
+ \varepsilon_{\alpha \beta \gamma \delta}
\langle L_\alpha \lambda_{ij} \rangle \langle L_\beta L_\gamma L_\delta
\lambda_{kl} \rangle \Big) \Big]           
\label{Eq:nlo2}
\end{eqnarray}
Comparing Eqs.(\ref{Eq:nlo1})-(\ref{Eq:nlo2}) with 
the  effective $\Delta S = 1$ weak chiral Lagrangian to order ${\cal O}(p^4)$ in
$\chi$PT given by
Ecker {\em et al.}~\cite{EKW} and Esposito-Far\`ese~\cite{Esposito}
(presented in Minkowski space):
\begin{eqnarray}
{\cal L}^{\Delta S=1,{\cal O}(p^4)}_{\rm eff}
&=& - {G_F \over \sqrt{2}} V_{\rm ud} V_{\rm us}^*
f_\pi^2 \left[ \left(N_{1}^{(\underline{8})}
        \langle \lambda_6  L_\mu  L^\mu  L_\nu L^\nu  \rangle
\; + \; N_{2}^{(\underline{8})} \cdot
        \langle \lambda_6  L_\mu  L^\nu  L_\nu L^\mu  \rangle
\right.\right. \nonumber \\
 && \hspace{2.5cm}+\; N_{3}^{(\underline{8})}
 \langle \lambda_6  L_\mu  L_\nu \rangle \langle  L^\mu L^\nu  \rangle
\;+ \; N_{4}^{(\underline{8})}
\langle \lambda_6  L_\mu \rangle \langle  L^\mu  L_\nu L^\nu  \rangle
\nonumber \\
&&\hspace{2.5cm}  +\; \left.N_{28}^{(\underline{8})}   i \,
        \epsilon_{\mu \nu \rho \delta}
        \langle \lambda_6  L^\mu \rangle
        \langle  L^\nu  L^\rho  L^\delta  \rangle \right)\nonumber \\
&+& \frac59 {t}_{ijkl} \, \left( N_{1}^{(\underline{27})}
        \langle \lambda_{ij} L_\mu L^\mu \rangle
        \langle \lambda_{kl} L_\nu L^\nu \rangle\right.
\;+\; N_{2}^{(\underline{27})}
       \langle \lambda_{ij} L_\mu L_\nu \rangle
        \langle \lambda_{kl} L^\mu L^\nu \rangle
\nonumber \\
&&\hspace{0.2cm} +\;  N_{3}^{(\underline{27})}
        \langle \lambda_{ij} L_\mu L_\nu \rangle
        \langle \lambda_{kl}  L^\nu L^\mu \rangle
\; +\; N_{4}^{(\underline{27})}
        \langle \lambda_{ij} L_\mu \rangle
        \langle \lambda_{kl} L_\nu L^\mu  L^\nu \rangle
\nonumber \\                                                         
&&\hspace{0.2cm} +\; N_{5}^{(\underline{27})}
    \langle \lambda_{ij} L_\mu \rangle
    \langle \lambda_{kl} \left\{ L^\mu ,  L_\nu  L^\nu \right\} \rangle
\;+\; N_{6}^{(\underline{27})}
        \langle L_\mu L^\mu \rangle
        \langle \lambda_{ij} L_\nu \rangle
        \langle \lambda_{kl} L^\nu \rangle
\nonumber \\
&& \hspace{0.2cm} \left.\left. +\; N_{20}^{(\underline{27})}
 i \,
        \epsilon_{\mu \nu \rho \delta}
        \langle \lambda_{ij} L^\mu L^\nu  \rangle
        \langle \lambda_{kl} L^\rho L^\delta \rangle
\;+\; N_{21}^{(\underline{27})}    i \,
        \epsilon_{\mu \nu \rho \delta}
        \langle \lambda_{ij} L^\mu \rangle
        \langle \lambda_{kl} L^\nu L^\rho L^\delta  \rangle
\right) \right] ,
\end{eqnarray}
we derive the LECs in the case of the leading order
in the large $N_c$ expansion:
\begin{eqnarray}
N_{1}^{(\underline{8})} &=&
\left( -{N_c^2 M^2 \over 128 \pi^4 f_\pi^2 } + {N_c \over 8 \pi^2}
- {f_\pi^2 \over 2 M^2} \right) c_6,
\\
N_{2}^{(\underline{8})} &=&
{N_c \over 60 \pi^2}
\Big( -2 c_1 + 3 c_2 + 5 c_4 - 3 c_9 + 2 c_{10} \Big),
\\
N_{3}^{(\underline{8})} &=&
0,
\\
N_{4}^{(\underline{8})} &=&
{N_c \over 60 \pi^2 }
\left( \frac{3}{2} c_1 - c_2 + \frac{5}{2} c_3 - \frac{5}{2} c_5
+ c_9 -  \frac{3}{2} c_{10} \right),
\\
N_{28}^{(\underline{8})} &=&
{N_c \over 60 \pi^2}
\left(  -\frac{3}{2} c_1 + c_2 - \frac{5}{2} c_3 - \frac{5}{2} c_5
- c_9 + \frac{3}{2} c_{10} \right),
\\                                                      
N_{1}^{(\underline{27})} &=&
N_{5}^{(\underline{27})} =
N_{6}^{(\underline{27})} =
N_{20}^{(\underline{27})} = 0,
\\
N_{2}^{(\underline{27})} &=&
-N_{3}^{(\underline{27})} =
-N_{4}^{(\underline{27})} =
N_{21}^{(\underline{27})} =
{N_c \over 60 \pi^2 }
\left( - 3 c_1 - 3 c_2 - \frac{9}{2} c_9 - \frac{9}{2} c_{10} \right).
\end{eqnarray}
As noted by G. Ecker {\em et al.}~\cite{EKW}, the LECs
$N_{1}^{(\underline{8})}$, $N_{2}^{(\underline{8})}$,
$N_{3}^{(\underline{8})}$ and $N_{4}^{(\underline{8})}$
contribute to the process $K\rightarrow 3\pi$ while
$N_{28}^{(\underline{8})}$ does to the radiative $K$-decays.  In particular,
the $N_{28}^{(\underline{8})}$ is related to the chiral
anomaly~\cite{BEP}.  The numerical results can be found in Table II.  Note
that the LECs in the eikosiheptaplet can assume only two values.

Taking into account the $1/N_c$ corrections, the LECs are extracted as follows:
\begin{eqnarray}
N_{1}^{(\underline{8})} &=&
\left( -{N_c^2 M^2 \over 128 \pi^4 f_\pi^2 } + {N_c \over 8 \pi^2}
- {f_\pi^2 \over 2 M^2} \right) c_6 \nonumber \\
&& +\;
\left( -{N_c M^2 \over 128 \pi^4 f_\pi^2 } + {1 \over 8 \pi^2}
- {f_\pi^2 \over 2 N_c M^2} 
\right) c_5,
\\
N_{2}^{(\underline{8})} &=&
{N_c \over 60 \pi^2}
\left( \left( -2 +{1 \over N_c} 3 \right) c_1 +
\left( 3 - {1 \over N_c} 2 \right) c_2
+ {1 \over N_c} 5 c_3 + 5 c_4 \right. \nonumber \\
&& + \; \left.
\left( - 3 + {1 \over N_c} 2 \right) c_9 +
\left( 2 - {1 \over N_c} 3 \right) c_{10} \right),
\\
N_{3}^{(\underline{8})} &=&
0,
\\                                                                          
N_{4}^{(\underline{8})} &=&
{N_c \over 60 \pi^2 }
\left( \left( \frac{3}{2} - {1 \over N_c} \right) c_1
+ \left( - 1 + {1 \over N_c} \frac{3}{2} \right) c_2
+ \frac{5}{2} c_3 + {1 \over N_c} \frac{5}{2} c_4
\right. \nonumber \\ && \hspace{10mm} \left.
- \frac{5}{2} c_5 - {1 \over N_c} \frac{5}{2} c_6
+ \left( 1 - {1 \over N_c} \frac{3}{2} \right) c_9
+ \left( -  \frac{3}{2} + {1 \over N_c} \right) c_{10} \right),
\\
N_{28}^{(\underline{8})} &=&
{N_c \over 60 \pi^2}
\left(  \left( -\frac{3}{2} + {1 \over N_c} \right) c_1
+ \left( 1 - {1 \over N_c}  \frac{3}{2} \right)  c_2
- \frac{5}{2} c_3 -  {1 \over N_c} \frac{5}{2} c_4
\right. \nonumber \\ && \hspace{10mm} \left.
- \frac{5}{2} c_5 - {1 \over N_c} \frac{5}{2} c_6
+ \left( - 1 + {1 \over N_c} \frac{3}{2} \right) c_9
+ \left(  \frac{3}{2} - {1 \over N_c} \right) c_{10} \right),
\\
N_{1}^{(\underline{27})} &=&
N_{5}^{(\underline{27})} =
N_{6}^{(\underline{27})} =
N_{20}^{(\underline{27})} = 0,
\\
N_{2}^{(\underline{27})} &=&
-N_{3}^{(\underline{27})} =
-N_{4}^{(\underline{27})} =
N_{21}^{(\underline{27})} =
{N_c \over 60 \pi^2 } \left( 1 + {1 \over N_c} \right)
\left( - 3 c_1 - 3 c_2 - \frac{9}{2} c_9 - \frac{9}{2} c_{10} \right).
\end{eqnarray}
Table III shows the effect of the ${\cal O}(N_c)$ corrections.
                                                                        
The effective $\Delta S = 2$ weak chiral Lagrangian is obtained as
\begin{eqnarray}
{\cal L}^{\Delta S=2,{\cal O}(p^4)}_{\rm eff}
&=& -\frac{G^2_{F} M^2_{W}}{4\pi^2}
{\cal F} \left(\lambda_c, \lambda_t, m^2_{c},m^2_{t},M^2_{W}\right)
b(\mu) \frac{N_c f^2_{\pi}}{12\pi^2}
\left(\langle \lambda_6 L_\mu L_\nu\rangle\langle\lambda_6 L^\nu L^\mu\rangle
\right.\nonumber \\ && -
\langle \lambda_6 L_\mu L_\nu\rangle\langle \lambda_6 L^\mu L_\nu\rangle
+ \langle \lambda_6 L_\mu \rangle\langle \lambda_6 L_\nu L^\mu L^\nu\rangle
- i\epsilon_{\mu\nu\alpha\beta}\langle \lambda_6 L^\mu
\rangle\langle \lambda_6 L^\nu L^\alpha L^\beta \rangle \Big).
\end{eqnarray}                       
The above formulae present the final result for the effective weak chiral 
Lagrangian for $\Delta S=1$ and $\Delta S=2$. It is constructed in a way 
to be used in $\chi$PT without further treatment, 
since all LECs are explicitely given. The result is a strict outcome of the
$\chi$QM and the expansion of its Lagrangian in powers of the momentum 
and in the number of colors. It is assumed, however, that no external 
fields are present.

\section{Summary and Conclusions}
The aim of the present work has been to derive the effective 
$\Delta
S=1,2$ effective weak chiral Lagrangian with its low energy 
constants from the chiral quark model using the weak effective action 
of Buchalla, Buras and Lautenbacher~\cite{Burasetal}. In leading order 
in the large $N_c$ expansion the Lagrangian is already known. The final 
result of our investigations is the effective weak chiral Lagrangian in 
next-to-leading order in $N_c$ and to fourth order in the momentum.

 As is already known, the contribution of leading order in the 
large $N_c$ expansion to the ratio $g_{\underline{8}}/g_{\underline{27}}$
is heavily underestimated in this model.  As we show in the present paper 
the inclusion of the next-to-leading order in
the large $N_c$ expansion does not help to improve this result. Hence 
on the present level of formalism and without further improvements the 
chiral quark model does not provide low energy constants which can 
directly be used in chiral perturbation theory for weak processes. Of course such a 
conclusion can finally only be drawn if a few actual observables have been 
calculated in chiral perturbation theory by using the above effective 
weak chiral Lagrangian. However, since the chiral quark model fails heavily 
in reproducing the ratio $g_{\underline{8}}/g_{\underline{27}}$
we do not have much hope that this will work.

Actually Antonelli {\em et al.}~\cite{Antonellietal}
have added to the lowest order in $N_c$ certain corrections from the gluon 
condensate 
known as of order ${\cal O} (\alpha_s N_c)$~\cite{PichRafael} in order to
change the ratio $g_{\underline{8}}/g_{\underline{27}}$. They have 
shown that the ${\cal O} (\alpha_s N_c)$ corrections indeed improve 
numerically the $\Delta T=1/2$ enhancement.  However, in our view 
there is an important caveat.  In fact, the ${\cal O} (\alpha_s N_c)^2$ 
is of the same order as ${\cal O} (\alpha_s N_c)$. The latter corrections 
were neglected by the authors of 
Ref.~\cite{PichRafael} hoping that they might be smaller
since they involve condensates of higher dimensions. In view of the large 
size of those corrections such an argument requires further 
substantiation, even though the numerical results are improved. In 
this paper no attempt was done to obtain gluonic corrections to the 
next-to-leading order in $N_c$.

At the present level of investigation we see the following ways of 
investigations, which might improve the low energy constants of the 
chiral quark model.

 First: The chiral quark model has been derived from QCD 
by Diakonov and Petrov~\cite{DP1,DP2,DP3,DP4} by 
assuming a gluonic vacuum configuration which 
consists of a dilute gas of interacting instantons and 
anti--instantons. As a result of this approach the constituent quark mass in
the chiral quark model is momentum-dependent and it is only an 
approximation to replace this by a regularization prescription with a 
properly chosen cut--off parameter. 
Thus it is interesting to investigate how far the present results change 
if such a momentum-dependent constituent mass is used. Such an 
investigation is even necessary if one wants to exploit fully the 
chiral quark model.

Second: All the results in the present paper are based on the assumption 
that the effective weak Hamiltonian of Buchalla, Buras and 
Lautenbacher can be used in connection with the chiral quark model.  This, 
however, is not that clear. If one considers the derivation of the chiral 
quark model from QCD by Diakonv and Petrov the renormalization point 
of the model is around 600 MeV corresponding to the average size 
of the instantons of 0.3 fm and the average distance of 1 fm. The 
Wilson coefficients of the effective weak Hamiltonian are 
evaluated at a scale of 1 GeV and it is not obvious if they 
can be used without further change at 600 MeV. Suggestions for 
investigations in this direction have recently been given in 
 \cite{NPB469,Bertolinietal,BijnensPrades}.

Actually in our next investigations we will follow the first suggestion and 
will incorporate the momentum-dependent quark mass in the 
chiral quark model. Such a procedure links the Lagrangian of the chiral 
quark model to QCD and, perhaps, the results will be improved.     

\section*{Acknowledgment} 
Authors thank M.V. Polyakov for valuable discussions and comments on
the present work.  The work is supported in part by COSY, DFG, 
and BMBF.  HCK wishes to acknowledge the financial support of 
the Korea Research Foundation made in the program year of 
1998.  KG thanks H. Toki and the RCNP (Osaka) for hospitality.

\newpage

\centerline{\large \bf Figures}

\vspace{0.8cm}

\centerline{\epsfig{file=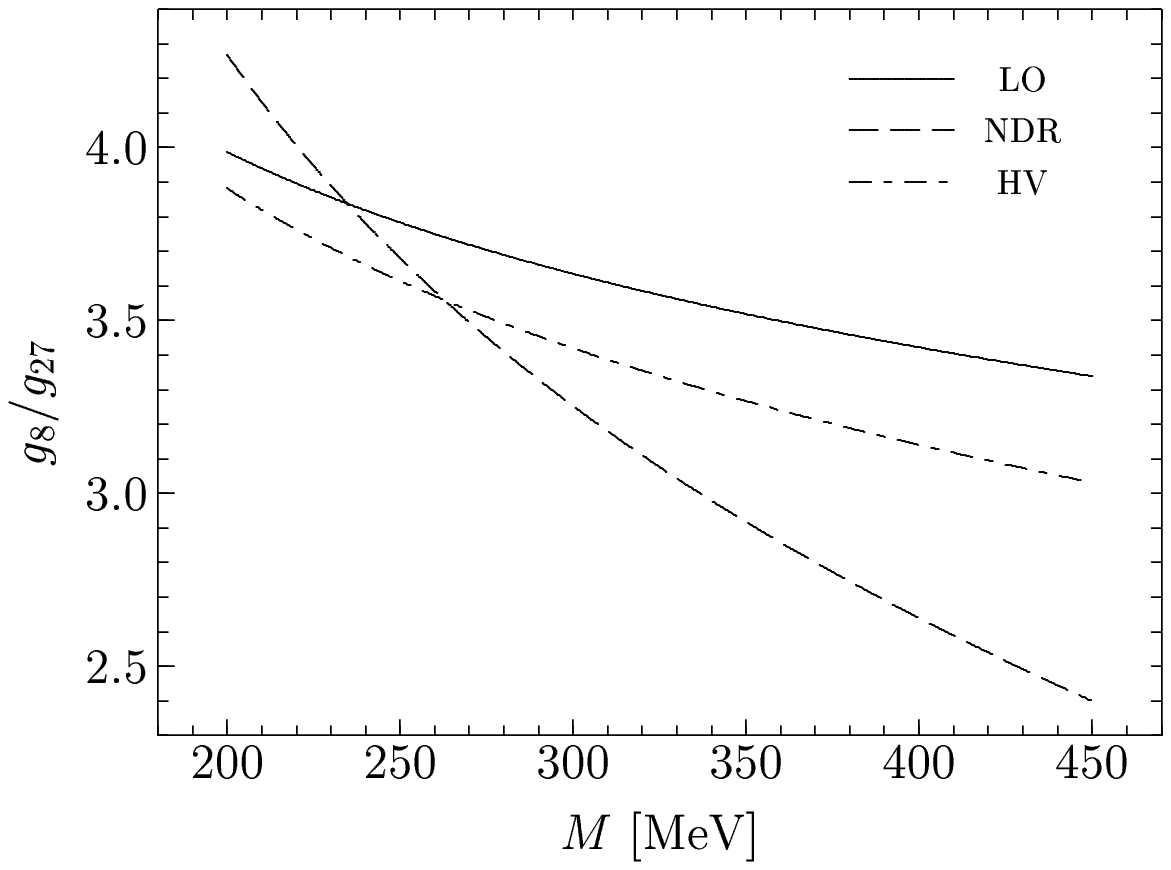,height=12.0cm,width=15.0cm,silent=}}

\vspace{0.8cm}
\noindent {\bf Fig.1}: Dependence of the 
$g_{\underline{8}}/g_{\underline{27}}$ on the 
$M$.  The solid curve denotes the LO renormalization scheme in 
Ref.\protect{\cite{Burasetal}}, while the dashed curve and dot-dashed one
stand for the NDR and the HV schemes, respectively.  The value of
the quark condensate $\langle \bar{q}q\rangle/2 =-(250\; {\rm MeV})^3$ 
is used.

\vfill
\newpage

\centerline{\epsfig{file=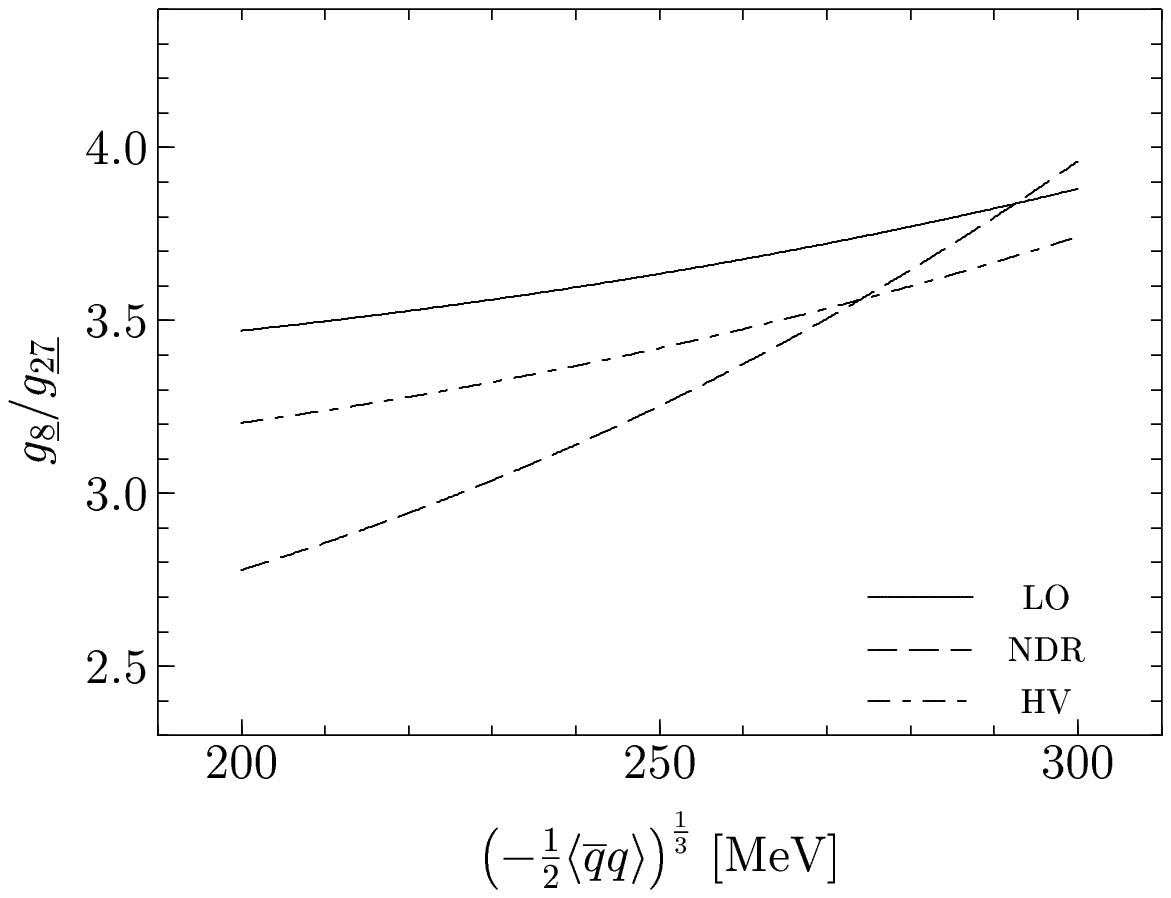,height=12.0cm,width=15.0cm,silent=}}

\vspace{0.8cm}
\noindent {\bf Fig.2}: Dependence of the 
$g_{\underline{8}}/g_{\underline{27}}$ on the $\langle 
\bar{q}q\rangle$.  The solid curve denotes the LO renormalization scheme in 
Ref.\protect{\cite{Burasetal}}, while the dashed curve and dot-dashed one
stand for the NDR and the HV schemes, respectively.  The value of
the constituent quark mass $M =300 \; {\rm MeV}$ is used.

\vfill
\newpage

\centerline{\epsfig{file=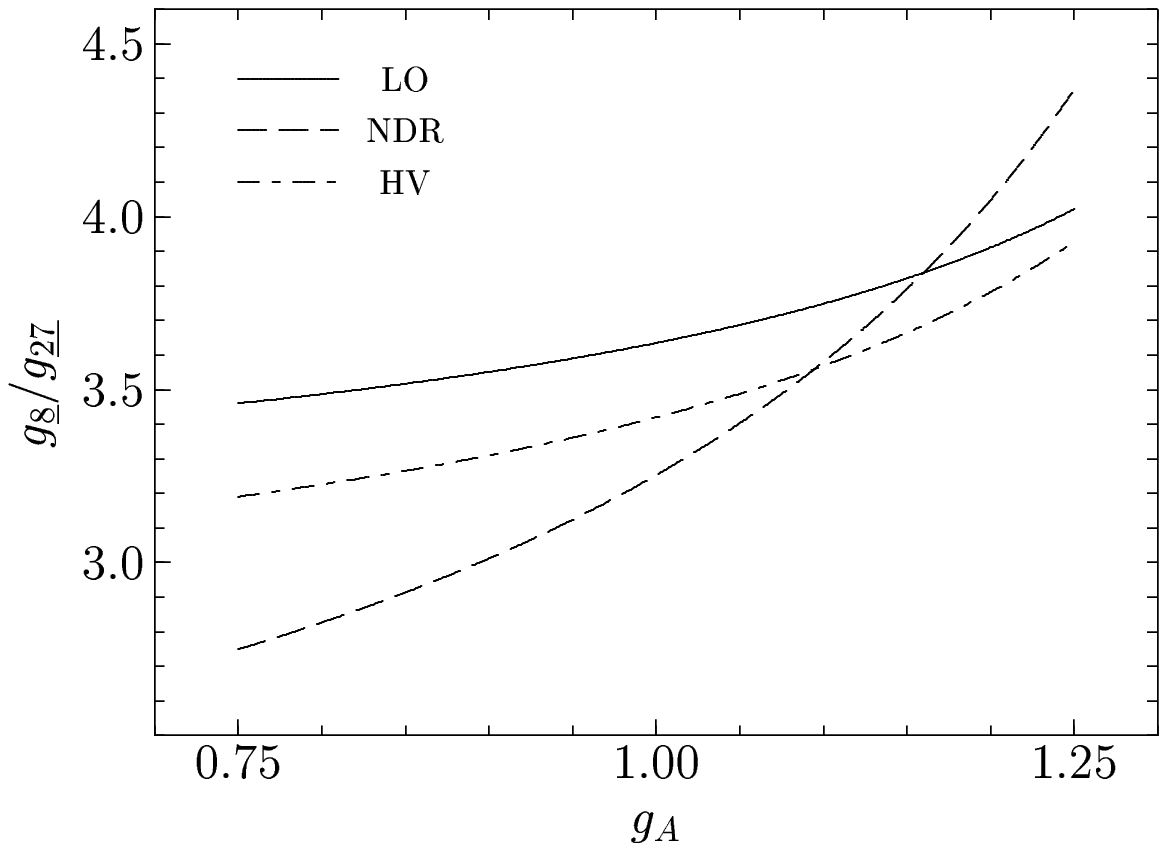,height=12.0cm,width=15.0cm,silent=}}

\vspace{0.8cm}
\noindent {\bf Fig.3}: Dependence of the 
$g_{\underline{8}}/g_{\underline{27}}$ on the quark axial-vector constant
$g_A$.  The solid curve denotes the LO renormalization scheme in 
Ref.\protect{\cite{Burasetal}}, while the dashed curve and dot-dashed one
stand for the NDR and the HV schemes, respectively.  The value of
the constituent quark mass $M =300 \; {\rm MeV}$ is used and the quark 
condensate $\langle \bar{q}q\rangle/2 =-(250\; {\rm MeV})^3$ 
is employed.

\vfill
\newpage

\begin{table}[th]
\caption{Wilson coefficients at $\mu=1 \; {\rm GeV}$. 
$c_i$ can be obtained by the relation $c_i (\mu) = z_i (\mu) + \tau y_i (\mu)$
which are provided by Ref.~\protect\cite{Burasetal}. }
\vspace{8pt}
\begin{math}
\begin{array}{|l|r|r|r||r|r|r||r|r|r|}
\hline
\hline
    & \multicolumn{3}{|c||}{\Lambda^{(4)}_{\overline{\rm MS}}=215\; {\rm MeV}}
    & \multicolumn{3}{|c||}{\Lambda^{(4)}_{\overline{\rm MS}}=325\; {\rm MeV}}
    & \multicolumn{3}{|c|}{\Lambda^{(4)}_{\overline{\rm MS}}=435\; {\rm MeV}}
    \cr \hline
    \multicolumn{1}{|c|}{\rm  Scheme} &
    \multicolumn{1}{|c|}{\rm  LO} &
    \multicolumn{1}{|c|}{\rm  NDR} &
    \multicolumn{1}{|c||}{\rm  HV} &
    \multicolumn{1}{|c|}{\rm  LO} &
    \multicolumn{1}{|c|}{\rm  NDR} &
    \multicolumn{1}{|c||}{\rm  HV} &
    \multicolumn{1}{|c|}{\rm  LO} &
    \multicolumn{1}{|c|}{\rm  NDR} &
    \multicolumn{1}{|c|}{\rm   HV}
    \cr \hline
c_1  & -0.607 & -0.409 & -0.494 & -0.748 & -0.509 & -0.640 & -0.907 & -0.625
   & -0.841 \\
c_2  &  1.333 &  1.212 &  1.267 &  1.433 &  1.278 &  1.371 &  1.552 &  1.361
   &  1.525 \\
c_3  &  0.003 &  0.008 &  0.004 &  0.004 &  0.013 &  0.007 &  0.006 &  0.023
   &  0.015 \\
c_4  & -0.008 & -0.022 & -0.010 & -0.012 & -0.035 & -0.017 & -0.017 & -0.058
   & -0.029 \\
c_5  &  0.003 &  0.006 &  0.003 &  0.004 &  0.008 &  0.004 &  0.005 &  0.009
   &  0.005 \\
c_6  & -0.009 & -0.022 & -0.009 & -0.013 & -0.035 & -0.014 & -0.018 & -0.059
   & -0.025 \\
\hline
c_7 /\alpha  &  0.004 &  0.003 & -0.003 &  0.008 &  0.011 & -0.002 &  0.011 & 
   0.021 & -0.001\\
c_8 /\alpha  & 0 &  0.008 &  0.006 &  0.001 &  0.014 &  0.010 &  0.001 & 
   0.027 &  0.017\\
c_9 /\alpha  &  0.006 &  0.008 &  0.001 &  0.009 &  0.019 &  0.006 &  0.013 & 
   0.035 &  0.012\\
c_{10} /\alpha  & 0 & -0.005 & -0.006 & -0.002 & -0.008 & -0.010 & -0.002
   & -0.015 & -0.018
 \cr \hline \hline
\end{array}
\end{math}
\end{table}

\begin{table}[th]
\caption{The low energy constants in ${\cal O}(N_c^{2})$ order.  The Wilson 
coefficients are from Ref.~\protect\cite{Burasetal} as shown in Table I.
$M= 300\,{\rm MeV}$ and $\langle \overline{q}q \rangle=-2 \cdot {(250\, {\rm MeV})}^3$
are used.}
\vspace{8pt}
\begin{math}
\begin{array}{|l|r|r|r||r|r|r||r|r|r|}
\hline
\hline
    & \multicolumn{3}{|c||}{\Lambda^{(4)}_{\overline{\rm MS}}=215\; {\rm MeV}}
    & \multicolumn{3}{|c||}{\Lambda^{(4)}_{\overline{\rm MS}}=325\; {\rm MeV}}
    & \multicolumn{3}{|c|}{\Lambda^{(4)}_{\overline{\rm MS}}=435\; {\rm MeV}}
    \cr \hline
    \multicolumn{1}{|c|}{\rm  Scheme} &
    \multicolumn{1}{|c|}{\rm  LO} &
    \multicolumn{1}{|c|}{\rm  NDR} &
    \multicolumn{1}{|c||}{\rm  HV} &
    \multicolumn{1}{|c|}{\rm  LO} &
    \multicolumn{1}{|c|}{\rm  NDR} &
    \multicolumn{1}{|c||}{\rm  HV} &
    \multicolumn{1}{|c|}{\rm  LO} &
    \multicolumn{1}{|c|}{\rm  NDR} &
    \multicolumn{1}{|c|}{\rm   HV}
    \cr \hline
g_{\underline 8} & 1.100 & 1.029 & 1.013 & 1.241 & 1.190 & 1.163 & 
  1.407 & 1.437 & 1.404 \\ 
 \hline
g_{{\underline{27}}} & 0.436 & 0.482 & 0.464 & 0.411 & 0.461 & 0.439 & 
  0.387 & 0.442 & 0.410 \\ 
 \hline
g_{\underline 8}/g_{{\underline{27}}} &  2.524 &  2.135 &  2.184 &  3.019 & 
   2.578 &  2.652 &  3.636 &  3.254 &  3.421 \\ 
 \hline
N_{1}^{(\underline{8})}  \cdot 10^3 & 0.16 & 0.39 & 0.16 & 0.23 & 0.61 
& 0.24 & 0.31 & 1.03 &  0.44 \\ 
N_{2}^{(\underline{8})}  \cdot 10^3 &26.21&22.01&24.01 &29.05&23.69 
& 26.89 & 32.35 & 25.54 & 30.96 \\ 
N_{3}^{(\underline{8})}  \cdot 10^3 & 0 & 0 & 0 & 0 & 0 & 0 & 0 & 0 & 0 \\ 
N_{4}^{(\underline{8})}  \cdot 10^3 &-11.37 &-9.22 &-10.16 &-12.94 & - 10.28  
&  -11.77 &  -14.74 &  -11.47 &-13.99 \\ 
N_{28}^{(\underline{8})} \cdot 10^3&11.29&9.07 &10.08&12.84 &10.08
&11.67&14.62 & 11.24 & 13.86\\ \hline
N_{1}^{(\underline{27})} \cdot 10^3 & 0 & 0 & 0 & 0 & 0 & 0 & 0 & 0 & 0 \\ 
N_{2}^{(\underline{27})} \cdot 10^3 &-11.03 &-12.20 &-11.75 &-10.41&-11.69 
&  -11.11 &  - 9.80 &  -11.19 &  -10.39 \\ 
N_{3}^{(\underline{27})} \cdot 10^3 &11.03 &12.20&11.75&10.41&11.69
& 11.11 &  9.80 & 11.19 & 10.39 \\ 
N_{4}^{(\underline{27})} \cdot 10^3 & 11.03 & 12.20 & 11.75 & 10.41
&11.69&11.11 &9.80&11.19 & 10.39 \\ 
N_{5}^{(\underline{27})} \cdot 10^3 & 0 & 0 & 0 & 0 & 0 & 0 & 0 & 0 & 0 \\ 
N_{6}^{(\underline{27})} \cdot 10^3 & 0 & 0 & 0 & 0 & 0 & 0 & 0 & 0 & 0 \\ 
N_{20}^{(\underline{27})} \cdot 10^3 & 0 & 0 & 0 & 0 & 0 & 0 & 0 & 0 & 0 \\ 
N_{21}^{(\underline{27})} \cdot 10^3 &  -11.03 &  -12.20 & - 11.75 & - 10.41 & 
  - 11.69 & - 11.11 & -  9.80 & - 11.19 & - 10.39 \\ 
\hline \hline 
\end{array}
\end{math}
\end{table}

\begin{table}[th]
\caption{Low energy constants with ${\cal O}(N^2_{c})$ and ${\cal O}(N_c)$ 
contributions.  The Wilson 
coefficients are from Ref.~\protect\cite{Burasetal} as shown in Table I.
$M= 300\,{\rm MeV}$ and $\langle \overline{q}q \rangle=-2 
\cdot {(250\, {\rm MeV})}^3$ are used.}
\vspace{8pt}
\begin{math}
\begin{array}{|l|r|r|r||r|r|r||r|r|r|}
\hline
\hline
    & \multicolumn{3}{|c||}{\Lambda^{(4)}_{\overline{\rm MS}}=215\; {\rm MeV}}
    & \multicolumn{3}{|c||}{\Lambda^{(4)}_{\overline{\rm MS}}=325\; {\rm MeV}}
    & \multicolumn{3}{|c|}{\Lambda^{(4)}_{\overline{\rm MS}}=435\; {\rm MeV}}
    \cr \hline
    \multicolumn{1}{|c|}{\rm  Scheme} &
    \multicolumn{1}{|c|}{\rm  LO} &
    \multicolumn{1}{|c|}{\rm  NDR} &
    \multicolumn{1}{|c||}{\rm  HV} &
    \multicolumn{1}{|c|}{\rm  LO} &
    \multicolumn{1}{|c|}{\rm  NDR} &
    \multicolumn{1}{|c||}{\rm  HV} &
    \multicolumn{1}{|c|}{\rm  LO} &
    \multicolumn{1}{|c|}{\rm  NDR} &
    \multicolumn{1}{|c|}{\rm   HV}
    \cr \hline
g_{\underline 8} &  0.794 &  0.773 &  0.739 &  0.892 &  0.902 &  0.845 &
   1.009 &  1.117 &  1.025 \\
 \hline
g_{{\underline{27}}} &  0.581 &  0.642 &  0.618 &  0.548 &  0.615 &  0.585 &
   0.516 &  0.589 &  0.547 \\
 \hline
g_{\underline 8}/g_{{\underline{27}}} &  1.368 &  1.204 &  1.196 &  1.628 &
   1.467 &  1.445 &  1.955 &  1.896 &  1.874 \\
 \hline
N_{1}^{(\underline{8})}  \cdot 10^{3} & 0.14 & 0.35 &   0.14 &   0.20 
&   0.57 & 0.22 & 0.29 & 0.98 &   0.41 \\
N_{2}^{(\underline{8})}   \cdot 10^{3} & 18.66 &15.91 &  17.26 &  
20.46 &  16.91 &19.08 &22.56 &17.98 &21.68 \\
N_{3}^{(\underline{8})}   \cdot 10^{3} & 0 & 0 & 0 & 0 & 0 & 0 & 0 & 0 & 0 \\
N_{4}^{(\underline{8})}   \cdot 10^{3} &  -6.96 &  -5.46 &  -6.12 
&  -8.05 &  -6.18&-7.23 &-9.28 &-6.96 &-8.72 \\
N_{28}^{(\underline{8})}  \cdot 10^{3} & 6.96 & 5.50 &   6.12 &
   8.05 &6.27 &7.25 & 9.30 &7.23 & 8.81 \\ \hline
N_{1}^{(\underline{27})}  \cdot 10^{3} & 0 & 0 & 0 & 0 & 0 & 0 & 0 & 0 & 0 \\
N_{2}^{(\underline{27})}  \cdot 10^{3}&-14.71&-16.27 &-15.66 & 
-13.88&-15.59&-14.81 & -13.07 & -14.92 & -13.86 \\
N_{3}^{(\underline{27})}  \cdot 10^{3} &14.71 &  16.27 &  15.66 & 
 13.88 &15.59 &14.81 &13.07 &14.92 &  13.86 \\
N_{4}^{(\underline{27})}  \cdot 10^{3} &  14.71 &  16.27 &  15.66 &
  13.88 &15.59 &14.81 &13.07 &14.92 & 13.86 \\
N_{5}^{(\underline{27})}  \cdot 10^{3} & 0 & 0 & 0 & 0 & 0 & 0 & 0 & 0 & 0 \\
N_{6}^{(\underline{27})}  \cdot 10^{3} & 0 & 0 & 0 & 0 & 0 & 0 & 0 & 0 & 0 \\
N_{20}^{(\underline{27})}\cdot 10^{3} & 0 & 0 & 0 & 0 & 0 & 0 & 0 & 0 & 0   \\
N_{21}^{(\underline{27})}\cdot 10^{3} & -14.71 & -16.27 & -15.66 & -13.88
   & -15.59 & -14.81 & -13.07 & -14.92 & -13.86 \\
\hline \hline 
\end{array}
\end{math}
\end{table}


\begin{thebibliography}{99}
\bibitem{GaillardLee} M.K. Gaillard and B.W. Lee, {\em Phys. Rev. Lett.}
{\bf 33} (1974) 108.
\bibitem{AM} G. Altarelli and L. Maiani, {\em Phys. Lett.}
{\bf B52} (1974) 351.
\bibitem{Witten} E. Witten, {\em Nucl. Phys.} {\bf B122} (1977) 109.
\bibitem{VZS} A.I. Vainshtein, V.I. Zakharov, and M. Shifman,
{\em JETP} {\bf 45} (1977) 670.
\bibitem{SVZ} M. Shifman, A.I. Vainshtein, and V.I. Zakharov,
{\em Nucl. Phys.} {\bf B120} (1977) 316.
\bibitem{GW} F.J. Gilman and M.B. Wise, {\em Phys. Rev.} {\bf D20}
(1979) 2392; {\em ibid.} {\bf D21} (1980) 3150.
\bibitem{GP} B. Guberina and R.D. Peccei, {\em Nucl. Phys.}
{\bf B163} (1980) 289.
\bibitem{BW} J. Bijnens and M.B. Wise, {\em Phys. Lett.} {\bf B137}
(1984) 245.
\bibitem{Burasetal} G. Buchalla, A.J. Buras, M.E. Lautenbacher,
{\em Rev. Mod. Phys.} {\bf 68} (1996) 1125 (and references therein).   
\bibitem{tHooft} G. 't Hooft, {\em Phys. Rev.} {\bf D14} (1976) 3432;
{\bf D18} (1978) 2199.
\bibitem{Witten2} E. Witten, {\em Nucl. Phys.} {\bf B156} (1979) 269.     
\bibitem{FYSY} M. Fukugita, T. Inami, N. Sakai, and S. Yasaki,
{\em Phys. Lett.} {\bf B72} (1977) 237.
\bibitem{NV} H.P. Nilles and V. Visnijc, {\em Phys. Rev.} {\bf D 19}
(1979) 969.
\bibitem{TT} D. Tadic and J. Trampetic, {\em Phys. Lett.} {\bf B114}
(1982) 179.  
\bibitem{CFG} R.S. Chivukula, J.M. Flynn, and H. Georgi,
{\em Phys. Lett.} {\bf B171} (1986) 453.        
\bibitem{GasserLeutwyler} J. Gasser and H. Leutwyler, {\em Ann. Phys.}
(N.Y.) {\bf 158} (1984) 142.    
\bibitem{Kamboretal} J. Kambor, J. Missimer, and D. Wyler,
{\em Nucl. Phys.} {\bf B346} (1990) 17; {\em Phys. Lett.}
{\bf 261B} (1991) 496.
\bibitem{Esposito} G. Esposito-Far\`ese, {\em Zeit. f. Phys.}
{\bf C50} (1991) 255.         
\bibitem{EKW} G. Ecker, J. Kambor, and D. Wyler, {\em Nucl. Phys.}
{\bf B394} (1993)101.   
\bibitem{DPP}D. Diakonov, V. Petrov and P. Pobylitsa,
{\em Nucl. Phys.} {\bf B272} (1988) 809  
\bibitem{BJW} A.J. Buras, M. Jamin, and P.H. Weisz, {\em Nucl. Phys.}
{\bf B347} (1990) 491.
\bibitem{HN1} S. Herrlich and U. Nierste, {\em Nucl. Phys.} {\bf B419}
(1994) 292; {\em ibid.} {\bf B476} (1996) 27.
\bibitem{HN2} S. Herrlich and U. Nierste, {\em Phys. Rev.} {\bf D52}
(1995) 6505.
\bibitem{Antonellietal}  V. Antonelli, S. Bertolini, J.O. Eeg,
M. Fabbrichesi and E.I. Lashin, {\em Nucl. Phys.} {\bf B469} (1996) 143.
\bibitem{Bertolinietal} S. Bertolini, J.O. Eeg, M. Fabbrichesi, and
E.I. Lashin, {\em Nucl. Phys.} {\bf B514} (1996) 63, 93. 
\bibitem{DE} D. Diakonov and M. Eides, {\em JETP Lett.}
{\bf 38} (1983) 433.
\bibitem{AF} J.R. Aitchison and C. Frazer,
{\em Phys. Lett.} {\bf 146B} (1984) 63;
{\em Phys. Rev.} {\bf D31} (1985) 2608.
\bibitem{chan} L.-H. Chan, {\em Phys. Rev. Lett.}
{\bf 55} (1985) 21.
\bibitem{dsw} A. Dhar, R. Shankar, and S. Wadia, {\em Phys. Rev.}
{\bf D31} (1985) 3256.      
\bibitem{Zuk} J.A. Juk, {\em Zeit. Phys.} {\bf C29} (1985) 21.
\bibitem{ERT} D. Espiru, E. de Rafael, and J. Taron,
{\em Nucl. Phys.} {\bf B345} (1990) 22.
\bibitem{Bijnens} J. Bijnens, {\em Phys. Rep.} {\bf 265} (1996) 369.  
\bibitem{WZ} J. Wess and B. Zumino, {\em Phys. Lett.} {\bf B37} (1971) 95.
\bibitem{Witten3} E. Witten, {\em Nucl. Phys.} {\bf B223} (1983) 422, 433.
\bibitem{GilmanWise2} F.J. Gilman and M. Wise, {\em Phys. Rev.}
{\bf D27} (1983) 1128.   
\bibitem{InamiLim} T. Inami and C.S. Lim, {\em Prog. Theor.Phys.}
{\bf 65} (1981) 297.   
\bibitem{BBH} G. Buchalla, A.J. Buras, and M.K. Harlander,
{\em Nucl. Phys.} {\bf B337} (1990) 313.        
\bibitem{Cronin} J.A. Cronin, {\em Phys. Rev.} {\bf 161} (1967) 1483.     
\bibitem{HV} G. 't Hooft and M. Veltman, {\em Nucl. Phys.}
{\bf B44} (1972) 189.
\bibitem{BM} P. Breitenlohner and D. Maison, {\em Comm. Math. Phys.}
{\bf 52} (1977) 11,39, 55.     
\bibitem{review} Ch. Christov, A. Blotz, H.-Ch. Kim, P. Pobylitsa,
T. Watabe, Th. Meissner, E. Ruiz Arriola, and K. Goeke,
{\em Prog. Part. Nucl. Phys.} {\bf 37} (1996) 91.        
\bibitem{BPKG} W. Broniowski, M.V. Polyakov, H.-Ch. Kim, and K. Goeke,
{\em Phys. Lett.} {\bf B438} (1998) 242.  
\bibitem{JJS} P. Jain, R. Johnson, and J. Schechter, {\em Phys. Rev.}
{\bf D38} (1988) 1571.    
\bibitem{Prasz} M. Prasza\l owicz, {\em Phys. Rev.} {\bf D42} (1990) 216.    
\bibitem{Weinberg} S. Weinberg, {\em Phys. Rev. Lett.}
{\bf 65} (1990) 1181; {\em Phys. Rev. Lett.} {\bf 67} (1990) 3473.
\bibitem{Dicusetal}D.A. Dicus, D. Minic, U. van Kolck, and R. Vega,
{\em Phys. Lett.} {\bf B284} (1992) 384.
\bibitem{BSL} W. Broniowski, A. Steiner, and M. Lutz , 
{\em Phys. Rev. Lett.} {\bf 71} (1993) 1787.    
\bibitem{PichRafael} A. Pich and E. de Rafael, {\em Nucl. Phys.}
{\bf B358} (1991) 311.
\bibitem{DP1} D. Diakonov and V. Petrov, {\em Phys. Lett.} {\bf B147} 
(1984) 351.
\bibitem{DP2} D. Diakonov and V. Petrov, {\em Nucl. Phys.} {\bf B272}
(1986) 457.
\bibitem{DP3} D. Diakonov and V. Petrov, ``{\em Spontaneous breaking 
of chiral symmetry in the instanton vacuum}'',
in Hadron matter under extreme conditions, Kiew (1986) p. 192.
\bibitem{DP4} D. Diakonov and V. Petrov, ``{\em Quark cluster dynamics}'',
Lecture Notes in Physics, Springer, Berlin (1992) 288.
\bibitem{BEP} J. Bijnens, G. Ecker, and A. Pich, {\em Phys. Lett.}
{\bf B286} (1992) 341.
\bibitem{NPB469} V. Antonelli, S. Bertolini, M. Fabbrichesi, and E.I. Lashin,
{\em Nucl. Phys.} {\bf B469} (1996) 181.
\bibitem{BijnensPrades} J. Bijnens and J. Prades, hep-ph/9811472, (1998).    

\end{thebibliography}
\end{document}